\documentclass[a4paper,11pt]{article}

\usepackage{bbm}
\usepackage{latexsym}
\usepackage{graphicx}
\usepackage{psfrag}
\usepackage{verbatim}

\newcommand{\be}{\begin{equation}}
\newcommand{\ee}{\end{equation}}
\newcommand{\bea}{\begin{eqnarray}}
\newcommand{\eea}{\end{eqnarray}}
\newcommand{\bean}{\begin{eqnarray*}}
\newcommand{\eean}{\end{eqnarray*}}

\renewcommand{\b}{\langle}
\renewcommand{\k}{\rangle}

\renewcommand{\i}{{\rm i}}
\newcommand{\e}{{\rm e}}

\renewcommand{\d}{{\rm d}}
\renewcommand{\c}[1]{{\cal #1}}

\newcommand{\pa}{\partial}

\renewcommand{\v}[1]{\underline{#1}} 

\renewcommand{\ss}{\scriptstyle}
\newcommand{\sss}{\scriptscriptstyle}
\newcommand{\ds}{\displaystyle}

\newcommand{\pdiff}[2]{\frac{\partial #1}{\partial #2}}

\newcommand{\bZ}{\mathbbm{Z}}

\newcommand{\bC}{\mathbbm{C}}
\newcommand{\bR}{\mathbbm{R}}

\newcommand{\G}{\c{G}}
\renewcommand{\H}{\c{H}}

\newcommand{\N}{\c{N}}
\renewcommand{\P}{\c{P}}

\newcommand{\T}{\c{T}}
\renewcommand{\P}{\c{P}}

\newcommand{\I}{\c{I}}

\newcommand{\eq}[1]{(\ref{#1})}

\newcommand{\fig}[1]{Fig.\ \ref{#1}}

\newcommand{\pic}[4]
{
 \begin{figure}
 \begin{center}
 \includegraphics[height=#3]{#4}
 \end{center}
 \caption{\label{#1} #2}
 \end{figure}
}

\newcommand{\hA}{\hat{A}}
\newcommand{\hE}{\hat{E}}
\newcommand{\he}{\hat{e}}
\newcommand{\hK}{\hat{K}}
\newcommand{\hH}{\hat{H}}
\newcommand{\hC}{\hat{C}}

\newcommand{\Ab}{\overline{A}} 

\newcommand{\St}{\tilde{S}} 
\newcommand{\gt}{\tilde{g}} 

\newcommand{\cb}{\overline{c}}
\newcommand{\htil}{\tilde{h}} 

\begin{document}
\thispagestyle{empty}
\hfill
\parbox[t]{3.6cm}{
hep-th/0409036 \\
AEI-2004-073} \vspace{2cm}

\begin{center}
{\bf\Large Free vacuum for loop quantum gravity}\\[4mm]
{Florian Conrady \\[2mm] 
\small\it Max-Planck-Institut f\"{u}r Gravitationsphysik, Albert-Einstein-Institut, D-14476 Golm, Germany\\
\small\it Centre de Physique Th\'eorique de Luminy, CNRS, F-13288,
France}
\end{center}
\vskip.5cm

\begin{abstract}
\noindent We linearize extended ADM-gravity around the flat torus, and use the associated Fock vacuum to construct a state that could play the role of a ``free'' vacuum in loop quantum gravity. The state we obtain is an element of the gauge-invariant kinematic Hilbert space and restricted to a cutoff graph, as a natural consequence of the momentum cutoff of the original Fock state. It has the form of a Gaussian superposition of spin networks. We show that the peak of the Gaussian lies at weave-like states and derive a relation between the coloring of the weaves and the cutoff scale. Our analysis indicates that the peak weaves become independent of the cutoff length when the latter is much smaller than the Planck length. By the same method, we also construct multiple-graviton states. We discuss the possible use of these states for deriving a perturbation series in loop quantum gravity.
\end{abstract}
\vskip1cm

\section{Introduction}

In the research on canonical loop quantum gravity (LQG) and spinfoam models\footnote{For an introduction, see e.g.\ \cite{T,R} and \cite{B,P}.}, one of the major open problems is the development of a reliable semiclassical analysis. At present, we do not know if these theories contain semiclassical states which correctly reproduce the observed Einstein gravity. Nor is a perturbative expansion available that would allow one to calculate the scattering of low-energy excitations of such states. Several lines of research have led to proposals for vacuum states, tentative results on perturbations, and modified dispersion relations for matter: among them are approaches based on the Kodama state \cite{S}-\cite{SoCh}, spin network invariants \cite{Mi1,Mi2}, linearized gravity \cite{ARSgravitonsloops,V}, coherent states \cite{TGCS1}-\cite{AL}, weaves \cite{ARSweaves}-\cite{CRe} and general boundaries \cite{CDORT}. While the kinematics of the gravitational sector and the dynamics of the matter sector are relatively well understood, the gravitational dynamics and its semiclassical limit remain largely unclear. Consequently, we have no conclusive evidence that loop quantum gravity has a physically realistic semiclassical and low-energy limit\footnote{For an overview on the literature, see sec.\ II.3, \cite{T}.}.

The work of this article is motivated by the following question: can one obtain a semiclassical perturbation series for LQG and, if yes, how? At first sight, it may seem odd to ask this question about a theory that has often been characterized as the {\it non-perturbative} alternative to perturbative approaches in quantum gravity. It is the very failure of traditional perturbation theory that led to the loop approach to quantum gravity, and one of its strongest points is the fact that it does not rely on any approximative scheme for its definition. On the other hand, we know that theories can have both perturbative and non-perturbative regimes, depending on which scale we look at them. In QED, for example, perturbative expansions work fine at low-energies where the fine-structure constant is small, while at higher energies the coupling running grows and perturbation theory breaks down. It is renormalization that relates the different regimes of the same theory, and allows us, in principle, to compute low-energy actions from actions at more fundamental scales. It gives us, in particular, the relation of measurable couplings at accessible scales to bare couplings at cutoff scales.

When the transition from one scale to another involves strong coupling regimes, the renormalization procedure may require non-perturbative techniques. Thus, it can happen that a field theory is perturbative at low energies, while a {\it non-perturbative} renormalization is needed to compute its low-energy action. This suggests that the failure of perturbative gravity may not come from the coupling expansion itself, but from the {\it perturbative} renormalization that fails to provide us with unique couplings in the effective action. Such a viewpoint is supported by the work of Reuter and others (see e.g.\ \cite{Reu}-\cite{PercPeri}) who investigate non-perturbative renormalization group flows of gravitational actions.

From that perspective, it is conceivable that loop quantum gravity is a non-perturbative theory at its fundamental scale, and nevertheless accessible to perturbative treatments at lower scales. While possible in principle, this scenario remains rather elusive, since so far the loop formalism does not dispose of the techniques to implement renormalization and perturbation theory. Tentative ideas on renormalization have been formulated by Markopoulou and Oeckl \cite{Ma,O}, and there are first attempts to develop a perturbation theory around the Kodama state \cite{So,SoCh}. Starodubtsev constructs string-like excitations of this state \cite{St}, and Smolin has analyzed string perturbations \cite{Sstring} of causal spin networks \cite{MaScausalevolution,MaSlocalcausality}. 

Here, we will take a more conservative strategy and try to proceed in close analogy with ordinary quantum field theory. There, the working principle of perturbation theory can be roughly described as follows: we choose a classical background and consider only states which are semiclassically peaked around it. As a result, the Hamiltonian is dominated by the lowest order terms in the fluctuation --- the free part --- whereas higher orders can be treated as small corrections. The free Hamiltonian defines a linear system and provides a first approximation to the dynamics of the perturbative regime. Inspite of its simplicity, it is far from clear how this scheme should be transferred to the framework of loop quantum gravity: how can we generalize field-theoretic concepts to a theory, where the basic degrees of freedom are not fields, but labelled abstract graphs? What is the analogue of a field fluctuation in a space of networks? What tells us how to separate an operator on graphs into a ``free'' and ``interaction'' part?

In the present paper, we approach this problem by constructing candidate states for a free vacuum and free gravitons in loop quantum gravity. We analyze these states in the hope to gain information on how semiclassical properties manifest themselves in the loop framework and how this could be exploited to do perturbation theory.

Since we do not know how to linearize LQG itself, our approach is indirect: we know how to linearize ADM gravity, and we know that field and loop-like degrees of freedom are physically related --- after all they arise from a quantization of the same classical theory. We employ this relation to translate the free vacuum and free gravitons of ADM gravity into states of the loop representation.  
In contrast to earlier attempts in this direction\cite{ARSgravitonsloops,V}, we arrive at states in the Hilbert space of the {\it full non-linearized} theory. An important feature of the construction lies in the fact that it starts from momentum-regularized states and translates this property into a cutoff graph of the final loop states. Thus, they can be viewed as more or less coarse-grained states, depending on the value of the cutoff parameter.

Both the vacuum and gravitons take the form of Gaussian superpositions of spin networks whose graphs lie on the cutoff graph. We analyze the maximum of the Gaussian and find that spin networks at the peak have similar properties as weaves \cite{ARSweaves}: we determine their graphs and establish a relation between mean spin and cutoff scale. Our analysis indicates that in the limit of very small cutoff length, the peak spin networks become independent of the cutoff graph, have spin label 1/2 and graphs of a length scale close to the Planck length. That is, the graphs of these spin networks maintain an effective Planck scale discreteness, while the mesh of the cutoff graph becomes infinitely fine.

The paper is organized as follows: in section \ref{Vacuum_of_linearized_Ashtekar_Barbero_gravity}, 
we linearize extended ADM gravity on a flat torus, quantize the reduced system and implement the 
linearized transformation to Ashtekar-Barbero variables. In this way, we obtain a free vacuum that is a 
functional of reduced triads or connections. Section \ref{Transition_to_degrees_of_freedom_of_LQG}
describes in several steps how we adapt this state to the degrees of freedom of loop quantum gravity. 
In section \ref{Graviton_states}, we use the same procedure to define loop analogues of many-graviton states.
Section \ref{Semiclassical_properties_of_the_vacuum_state} describes the peak analysis.
In the final section, the construction and properties of the states are summarized, and we discuss the similarities and differences to other proposals for vacuum states. We also mention ideas on a genuine loop quantization of the free vacuum which would allow one to remove the cutoff graph similarly as in the definition of Hamiltonian constraint and area operator. In the last subsection, we return to the issue of perturbation theory: we discuss how our states might be used for extracting a free part of the Hamiltonian constraint, and how they could be applied in renormalization. Interestingly, we arrive at an ansatz that is closely related to Smolin's string perturbations \cite{Sstring}.

\subsubsection*{\it Notation and conventions}

\renewcommand{\arraystretch}{1.5}
\begin{tabular}{ll}
\hline
spatial indices: $a,b,c \ldots = 1,2,3$ & LQG: loop quantum gravity  \\
internal indices: $i,j,k \ldots = 1,2,3$ & $S$: gauge-invariant spin network state  \\
3-metric: $g_{ab}$ & $\St$: gauge-{\it variant} spin network state  \\
determinant of 3-metric: $g$ & $\H$: kinematic Hilbert space of LQG  \\
Planck length: $l_p = \sqrt{\hbar\kappa}$ & \parbox[t]{8cm}{$\H_0$: gauge-invariant kinematic Hilbert space} \\ 
\parbox[t]{8cm}{gravitational coupling constant: $\kappa = 8\pi G / c^3$}  
& \raisebox{0cm}[0cm][0.75cm]{\parbox[t]{8cm}{$\H_{\rm diff}$: gauge- and 3d-diff-invariant \\ kinematic Hilbert space}} \\ \hline
\end{tabular}
\vspace{0.3cm} \renewcommand{\arraystretch}{1}

\noindent
We use units in which $c = 1$.

\section{Vacuum of linearized Ashtekar-Barbero gravity}
\label{Vacuum_of_linearized_Ashtekar_Barbero_gravity}

Loop quantum gravity is based on a quantization of the so-called Ashtekar-Barbero variables --- classical phase space variables that arise from a canonical transformation of the standard (extended) ADM variables. In order to arrive at some linearized version of LQG, we have, loosely speaking, six possibilities, corresponding to the different orders in which quantization ($Q$), linearization ($L$) and canonical transformation ($C$) can be applied. Let us abbreviate them by
\bean
1.\ LCQ && 4.\ QLC \\
2.\ CLQ && 5.\ QCL \\
3.\ LQC && 6.\ CQL\,,
\eean
where the order of operations goes from left to right, starting with the classical ADM-theory and ending up with a linearized form of LQG. Combinations $4.$ and $5.$ are merely hypothetical, since the ADM-variables have never been rigorously quantized. Ideally, what we want is number $6.$, a linearization of full LQG. So far, however, we do not know how to do this (or, for that matter, what linearization should exactly mean in that case), since the degrees of freedom of the theory are quite different from fields. Hence, before that problem is resolved, we have to content ourselves with options $1.$ to $3.$ Actually, $1.$ is identical to $2.$ (what other meaning should be given to $LC$ than $CL$?), so there remain possibilities $CLQ$ and $LQC$: the former has been explored by Ashtekar, Rovelli and Smolin \cite{ARSgravitonsloops} for imaginary, and by Varadarajan \cite{V} for real Immirzi parameter, when they applied a loop quantization to the linearized Ashtekar-Barbero variables. The route we follow in this paper takes the third variant $LQC$ as its point of departure: we linearize the classical extended ADM-gravity, apply a Schr{\"o}dinger quantization to it, and implement the linearized canonical transformation within the quantum theory. Thus, we arrive at a vacuum state which, at first glance, has little to do with loop-like degrees of freedom. The problem of relating this state to LQG will be the subject of section \ref{Transition_to_degrees_of_freedom_of_LQG}.

\subsection{Linearization of classical extended ADM formulation}
\label{Linearization_of_classical_extended_ADM_formulation}

In classical theory, linearization rests, similarly as in quantum theory, on the idea that we choose a background, restrict attention to small deviations from it, and exploit this to give a lowest order approximation for the dynamics. In the following, we go through the classcial linearization of extended ADM gravity, but we will not motivate or derive each step. The procedure is similar to that for standard ADM and complex Ashtekar gravity, which has been described in the literature\footnote{For a detailed exposition of linearization in the Hamiltonian context, we refer the reader to \cite{ALe}.}\cite{Hi,ARSgravitonsloops}.

To keep things as simple as possible, we choose space to be the 3-torus $T^3$ and linearize around a flat background on $T^3\times\bR$. The linearization consists of the following steps: we linearize the classical constraints, use them to obtain the reduced phase space, and determine the Poisson brackets and Hamiltonian on it. Once we have obtained the reduced classical system, we will quantize it using a Schr{\"o}dinger representation (sec.\ \ref{Reduced_phase_space_quantization}) and, finally, apply the linearized form of the canonical transformation (sec.\ \ref{Canonical_transformation}).

In the extended ADM formulation (see \cite{T}), the phase space variables are the 1-density triad
\be
E_i(\v{x}) = E^a_i(\v{x})\pdiff{}{x^a}\,,\quad i = 1,2,3,
\ee
and the canonically conjugate one-forms
\be
K^i(\v{x}) = K^i_a(\v{x})\d x^a\,,\quad i = 1,2,3,
\ee
They are relatted to the 3-metric $g_{ab}$ and extrinsic curvature $K_{ab}$ by
\bea
g_{ab} &=& |E|\,E^i_aE^i_b\,, \\
K^k_a &=& K_{ab}E^{bk}/\sqrt{g}\,,
\eea
where $E$ denotes the determinant $\det(E^a_i)$.
The Poisson brackets read
\be
\Big\{E^a_i(\v{x}),E^b_k(\v{y})\Big\} = \Big\{K^j_a(\v{x}),K^k_b(\v{y})\Big\} = 0\,,
\ee
\be
\Big\{E^a_i(\v{x}),K^k_b(\v{y})\Big\} = \frac{\kappa}{2}\,\delta^a_b\delta^k_i\,\delta(\v{x}-\v{y})\,.
\ee
The constraints are first-class and consist of the gauge, vector and Hamiltonian constraint:
\bea
G_{jk} &=& K_{a[j}E^a_{k]} \label{gaugeconstraint} \\
V_a &=& -D_b[K^j_a E^b_j - \delta^b_aK^j_cE^c_j] \\
C &=& \frac{1}{\sqrt{|E|}}\left(K^l_aK^j_b - K^j_aK^l_b\right)E^a_jE^b_l - \sqrt{|E|}\,R(E)\,.
\label{Hamiltonianconstraint}
\eea
$R(E)$ stands for the 3d Riemann tensor when written as a function of the densitized triad.

\subsubsection*{Linearization around flat torus}

Choose a flat classical background triad on $T^3$ such that the torus corresponds to a cube with macroscopic side length $L$ and periodic boundary conditions. Moreover, choose, once and for all, a coordinate system in which the background $E_i$- and $K^i$-fields read
\be
E^a_{cl\:k} = \delta^a_k\,,\quad K^k_{cl\:a} = K_{cl\:ab}E^{bk}_{cl}/\sqrt{g} = 0\,.
\ee
We introduce the relative variables
$$e^a_k := E^a_k - \delta^a_k\,,\qquad K^k_a = K^k_a - 0$$
and adher from now on to the convention that
\be
e_{ka} \equiv e^a_k\,,\quad K^{ka} \equiv K^k_a\,.
\ee  
That is, the spatial index of $e$ can be freely moved between upper and lower right position, while the spatial index of $K$ may be either in the lower or upper right position. The Poisson brackets of the relative variables are 
\be
\Big\{e_{ab}(\v{x}),K^{cd}(\v{y})\Big\} 
= \Big\{E^b_a(\v{x})-\delta^b_a,K^c_d(\v{y})\Big\}
= \frac{\kappa}{2}\,\delta^c_a\delta^d_b\,\delta(\v{x}-\v{y})\,.
\ee
By keeping only linear terms in $e$ and $K$, we arrive at the linearized constraints
\bea
G^{ab} &=& K^{[ab]}\,, \\
V^a &=& -\pa_b K^{ba} + \pa^a K\,, \\
C &=& 2\pa^a\pa^b e_{ab}\,,
\eea
which are again first-class. 

For the phase space reduction, it is convenient to change to Fourier space. On the 3-torus, we use the following conventions for Fourier series:
\bea
f(\v{x}) &=& \frac{1}{\sqrt{V}}\sum_{\v{k}}\e^{\i\v{k}\cdot\v{x}}\,f(\v{k})\quad\mbox{and} \\
f(\v{k}) &=& \frac{1}{\sqrt{V}}\int\d^3 x\;\e^{-\i\v{k}\cdot\v{x}}\,f(\v{x})\,,
\eea
where the wavevector $\v{k}$ takes values in $2\pi/L\,\bZ^3$. The delta functions on position and Fourier space have the respective transforms
\be
\delta(\v{x}-\v{x}') = \frac{1}{V}\sum_{\v{k}}\e^{\i\v{k}\cdot(\v{x}-\v{x}')}\,.
\ee
and
\be
\delta_{\v{k},\v{k}'} = \frac{1}{V}\int\d^3 x\;\e^{-\i(\v{k}-\v{k}')\cdot\v{x}}\,.
\ee
The Poisson bracket becomes
\be
\Big\{e_{ab}(\v{k}),K^{cd*}(\v{k}')\Big\} = \frac{\kappa}{2}\,\delta^c_a\delta^d_b\,\delta_{\v{k},\v{k}'}\,.
\ee
By imposing the linearized constraints and choosing suitable gauge conditions, we require that $e_{ab}$ and $K^{ab}$ are symmetric, transverse and have a constant trace. This defines our reduced phase space. We denote the reduced variables by $e^{\sss\rm red}_{ab}$ and $K^{ab}_{\sss\rm red}$, and write the Poisson bracket again as $\{\,,\,\}$. 

When Fourier-transformed, the reduced variables can be decomposed into six zero-mode components, and
two components for each nonzero $\v{k}$, corresponding to the two polarizations of gravitational waves:
\bea
e^{\sss\rm red}_{ab}(\v{x}) &=& \frac{1}{\sqrt{V}}\left(\sum_{i=1}^6\; e_i(0)\epsilon_{i\: {ab}}(0) +
\sum_{k > 0}\sum_{i=1}^2\;\e^{\i\v{k}\cdot\v{x}}\,e_i(\v{k})\epsilon_{i\: {ab}}(\v{k})\right)\,, \\
K^{ab}_{\sss\rm red}(\v{x}) &=& \frac{1}{\sqrt{V}}\left(\sum_{i=1}^6\; K_i(0)\epsilon_{i\: {ab}}(0) +
\sum_{k > 0}\sum_{i=1}^2\;\e^{\i\v{k}\cdot\v{x}}\,K_i(\v{k})\epsilon_{i\: {ab}}(\v{k})\right)\,.
\eea
More specifically, for each nonzero pair $\{\v{k},-\v{k}\}$ we choose a right-handed coordinate system s.t.\ one of the vectors, say $\v{k}$, points in the positive 3-direction. Then, we define the polarization tensors by
\bea
\epsilon_{1\: ab}(\v{k}) &:=& \frac{\i}{\sqrt{2}}\left(\delta_{1a}\delta_{2b} + \delta_{2a}\delta_{1b}\right)\,, \\
\epsilon_{2\: ab}(\v{k}) &:=& \frac{1}{\sqrt{2}}\left(\delta_{1a}\delta_{1b} - \delta_{2a}\delta_{2b}\right)\,, \\
\epsilon_{1\: ab}(-\v{k}) &:=& -\epsilon_{1\: ab}(\v{k})\,, \\
\epsilon_{2\: ab}(-\v{k}) &:=& \epsilon_{2\: ab}(\v{k})\,.
\eea
It follows that
\be
\epsilon^*_{i\:ab}(\v{k})\epsilon_{j\:ab}(\v{k}) = \delta_{ij}\,,
\ee
and 
\be
\epsilon^*_{i\:ab}(\v{k}) = \epsilon_{i\:ab}(-\v{k})\,.
\ee
For $k = 0$, we take $\epsilon_{i\: ab}(0)$, $i=1,\ldots,6,$ to be an orthonormal basis in the space of symmetric 2-tensors, i.e.
\be
\epsilon^*_{i\:ab}(0)\epsilon_{j\:ab}(0) = \delta_{ij}\,.
\ee
The projection of an arbitrary two-tensor $T_{ab}$ onto its reduced part is given by
\be
(PT)_{ab}(\v{x}) := \frac{1}{\sqrt{V}}\sum_{\v{k}}\;\e^{\i\v{k}\cdot\v{x}} P_{ab}{}^{cd}(\v{k})T_{cd}(\v{k})\,,
\ee
where 
\be
P_{ab}{}^{cd}(\v{k}) = \epsilon_{i\:ab}(\v{k})\epsilon^{cd*}_i(\v{k})\,.
\ee
Recall that the Poisson brackets of the reduced phase space are the pull-back of the Poisson brackets on the full phase space. In our notation,
\be
\Big\{e^{\sss\rm red}_{ab}(\v{x}),K^{cd}_{\sss\rm red}(\v{y})\Big\} = \Big\{(Pe)_{ab}(\v{x}),(PK)^{cd}(\v{y})\Big\}\,.
\ee
This implies that 
\bea
\Big\{e^{\sss\rm red}_{ab}(\v{k}),K^{cd*}_{\sss\rm red}(\v{k}')\Big\} &=& \Big\{(Pe)_{ab}(\v{k}),(PK)^{cd*}(\v{k}')\Big\} \nonumber \\
&=& P_{ab}{}^{ef}(\v{k}) P^{cd}{}_{gh}(\v{k}')\Big\{e_{ef}(\v{k}),K^{gh*}(\v{k}')\Big\} \nonumber \\
&=& P_{ab}{}^{ef}(\v{k}) P^{cd}{}_{gh}(\v{k}')\,\frac{\kappa}{2}\delta^g_e\delta^h_f \delta_{\v{k},\v{k}'} 
\nonumber \\
&=& \frac{\kappa}{2}P_{ab}{}^{cd}(\v{k})\delta_{\v{k},\v{k}'}\,.
\label{DiracbracketFourier}
\eea
Equivalently, we have
\be
\Big\{e_i(\v{k}),K^*_j(\v{k}')\Big\} = \frac{\kappa}{2}\delta_{ij}\delta_{\v{k},\v{k}'}
\label{Diracbracketpolarization}
\ee
for polarization and zero mode components.

Let us come to the linearized dynamics: the Hamiltonian is given by 
\be
H = \frac{1}{\kappa}\int\d^3 x\;N_{cl}(\v{x})C_{\rm quadr}(\v{x})
\label{definitionHamiltonian}
\ee
where $C_{\rm quadr}$ is the quadratic part of the Hamiltonian constraint \eq{Hamiltonianconstraint} when evaluated on the reduced phase space, and $N_{cl}$ is the lapse density associated to the background: i.e.\ the lapse for which $\int\d^3 x\;N(\v{x})C(\v{x})$ generates a flow that leaves the phase space point of the background fixed. In the case of the flat background, this is just $N_{cl} \equiv \sqrt{g}$. 
A straightforward calculation yields
\be
H = \frac{1}{\kappa}\int\d^3 x\;\left[K^{ab}_{\sss\rm red}(\v{x})K^{ab}_{\sss\rm red}(\v{x}) + \pa_c e^{\sss\rm red}_{ab}(\v{x})\pa_c e^{\sss\rm red}_{ab}(\v{x})\right]\,.
\ee
When expressed in terms of Fourier or polarization components, the Hamiltonian reads
\bea
H &=& \frac{1}{\kappa}\sum_{\v{k}} 
\left[K^{ab*}_{\sss\rm red}(\v{k})K^{ab}_{\sss\rm red}(\v{k}) 
+ k^2\,e^{\sss\rm red*}_{ab}(\v{k})e^{\sss\rm red}_{ab}(\v{k})\right] \\
&=& \frac{1}{\kappa}\sum_{\v{k}} 
\left[K^*_i(\v{k})K_i(\v{k}) + k^2\,e^*_i(\v{k})e_i(\v{k})\right]\,.
\eea
The polarization components for $\v{k}\neq 0$ describe the spatial change in $e^{\sss\rm red}_{ab}(\v{x})$: they oscillate in harmonic potentials and always stay near $e_i(\v{k}) = 0$. The zero modes are the constant part of the $e^{\sss\rm red}_{ab}(\v{x})$ field and move in a flat potential. This means that, to linear approximation, the overall shape of the torus behaves like a free particle. Unless the initial momentum is zero, the size of $e_i(0)$ will grow, so that at some point the linear approximation breaks down. This instability is due to the compactness of the torus. On $\bR^3$, the zero modes are absent and the linearization stable.

\subsection{Reduced phase space quantization}
\label{Reduced_phase_space_quantization}

We quantize the reduced system by introducing operators $\he^{\sss\rm red}_{ab}(\v{k})$ and $\hK^{ab}_{\sss\rm red}(\v{k})$, and replace the Poisson brackets \eq{Diracbracketpolarization}
and \eq{DiracbracketFourier} by commutation relations
\bea
\Big[\he^{\sss\rm red}_{ab}(\v{k}),\hK^{cd\dagger}_{\sss\rm red}(\v{k}')\Big] &=& \i\hbar\frac{\kappa}{2}P_{ab}{}^{cd}(\v{k})\delta_{\v{k},\v{k}'}\ \,,
\label{commutatorreducedFourier} \\
\Big[\he_i(\v{k}),\hK^\dagger_j(\v{k}')\Big] &=& \i\hbar\frac{\kappa}{2}\delta_{ij}\delta_{\v{k},\v{k}'}\,.
\label{commutatorpolarization}
\eea
The Hamilton operator becomes
\bea
\hH_{\rm quadr} &=& \frac{1}{\kappa}\sum_{\v{k}}\left[\hK^{ab\dagger}_{\sss\rm red}(\v{k})\hK^{ab}_{\sss\rm red}(\v{k}) 
+ k^2\,\he^{\sss\rm red\dagger}_{ab}(\v{k})\he^{\sss\rm red}_{ab}(\v{k})\right] \\
&=& \frac{1}{\kappa}\sum_{\v{k}} 
\left[\hK^\dagger_i(\v{k})\hK_i(\v{k}) + k^2\,\he^\dagger_i(\v{k})\he_i(\v{k})\right]\,.
\label{Hamiltonoperator}
\eea
To represent these operators, we use a Schr{\"o}dinger representation in terms of functionals of $e_i(\v{k})$, where $\he_i(\v{k})$ and $\hK^i(\v{k})$ act as multiplicative and derivative operators respectively:
\be
\he_i(\v{k}) \to e_i(\v{k})\,,\quad \hK_i(\v{k})\to 
-\i\hbar\frac{\kappa}{2}\pdiff{}{e^*_i(\v{k})}\,.
\ee
Note that the coefficients $e_i(\v{k})$ have to satisfy the reality condition $e^*_i(\v{k}) = e_i(-\v{k})$.
That is, $e_i(0)$ is real, and for $\v{k}\neq 0$ only half of the coefficients (say those for $k^1>0$) can be taken as independent variables. Thus, we choose the functional measure as 
\be
\int De := \left(\prod_{i=1}^6\int_{-\infty}^\infty\d e_i(0)\right)
\left(\prod_{\v{k}\neq 0,\:k^1>0}\;\prod_{i=1}^2\prod_{r=0}^1\int_{-\infty}^\infty\d e_{ir}(\v{k})\right)\,.
\ee
The additional index $r = 0,1,$ denotes the real and imaginary part respectively. 

Next we specify a free vacuum for the system. We require of it that it is time-independent and peaked around $e_i(\v{k}) = 0$. The peakedness is necesssary to ensure consistence with linearization\footnote{Of course, the peak property is just a minimum requirement: even then, the peak could be too wide, so that linearization is not applicable.}. At first, one might think that we look for the ground state of the Hamiltonian \eq{Hamiltonoperator}:
\be
\Psi_G[e_i(\v{k})] = \N\exp\left[-\frac{1}{\hbar\kappa}\sum_{\v{k}}\;k\,e^*_i(\v{k})e_i(\v{k})\right]\,.
\label{groundstate}
\ee
For nonzero $\v{k}\neq 0$, this functional is a Gaussian around $e_i(\v{k}) = 0$ and satisfies our requirements. A dependence on the zero modes is missing, however, so $\Psi_G$ has an infinite spread in $e_i(0)$, which is inconsistent with linearization. To remedy this, we add a Gaussian factor 
\be
\exp\left(-\frac{1}{\hbar\kappa}\,\omega_0\,e^2_i(0)\right)\,,\quad\omega_0 > 0\,,
\ee
for each zero mode. This gives us the new state
\be
\Psi[e_i(\v{k})] = \N\exp\left[-\frac{1}{\hbar\kappa}\sum_{\v{k}}\;\omega(\v{k})\,e^*_i(\v{k})e_i(\v{k})\right]\,,
\label{vacuum}
\ee
where 
\be
\omega(\v{k}) = 
\left\{
\parbox{2cm}{
$\begin{array}{ll}
k\,, & k > 0\,, \\
\omega_0\,, & k = 0\,.
\end{array}$
}
\right.
\ee
Note that under evolution by $\hH$, the spreading of the zero mode wavefunction proceeds on a time scale $\tau \sim 1 / \omega_0$. By choosing $\omega_0$ of the order $1/L$ or smaller, we can make the state $\Psi$ practically time-independent for all microscopic processes.

In the following, we use this state as the free vacuum of ADM gravity on $T^3\times\bR$. (Throughout the text we write normalization factors unspecifically as $\N$ and do not keep track of their precise value.)

\subsection{Canonical transformation}
\label{Canonical_transformation}

The classical Ashtekar-Barbero variables are obtained by the transformation 
\be
A^i_a = \Gamma^i_a + \beta K^i_a\,,
\label{canonicaltransformation}
\ee
where $\beta$ is the Immirzi parameter and $\Gamma^i_a[E]$ denotes the spin connection as a function of $E$:
\be
\Gamma^i_a[E] = \frac{1}{2}\varepsilon^{ijk}E^b_k\left[\pa_b E^j_a - \pa_a E^j_b + E^c_jE^l_a \pa_b E^l_c\right]
+ \frac{1}{4}\varepsilon^{ijk}\left[2 E^j_a\frac{\pa_b E}{E} - E^j_b\frac{\pa_a E}{E}\right]
\ee
Here, we take $\beta$ to be real. The transformation leads to the new Poisson brackets
\be
\Big\{E^a_i(\v{x}),A^j_b(\v{y})\Big\} 
= \Big\{E^a_i(\v{x}),\Gamma^j_b[E](\v{y}) + \beta K^j_b(\v{y})\Big\}
= \frac{\beta\kappa}{2}\,\delta^b_a\delta^j_i\,\delta(\v{x}-\v{y})\,,
\ee
or in terms of Fourier modes
\be
\Big\{E^a_i(\v{k}),A^{j*}_b(\v{k}')\Big\} = \frac{\beta\kappa}{2}\,\delta^b_a\delta^j_i\,\delta_{\v{k},\v{k}'}\,.
\label{newPoissonbracketFourier}
\ee
The linearization of \eq{canonicaltransformation} induces a canonical transformation on the reduced variables: \be
A^{ab}_{\sss\rm red} = \varepsilon_{acd}\pa_c e^{\sss\rm red}_{db} + \beta K^{ab}_{\sss\rm red}
\label{linearizedcanonicaltransformationclassical}
\ee 
One may check that $A^{ab}_{\sss\rm red}$ is again symmetric, transverse and of constant trace.

In the quantum theory, we introduce the new operator
\be
\hA^{ab}_{\sss\rm red} = \varepsilon_{acd}\pa_c \he^{\sss\rm red}_{db} + \beta \hK^{ab}_{\sss\rm red}\,.
\label{linearizedcanonicaltransformation}
\ee
Using that
\be
\i\varepsilon_{acd}k_c\epsilon_{1\,db} = k\epsilon_{2\,ab}\quad\mbox{and}\quad
\i\varepsilon_{acd}k_c\epsilon_{2\,db} = k\epsilon_{1\,ab}\,,
\ee
we see that \eq{linearizedcanonicaltransformation} is equivalent to
\renewcommand{\arraystretch}{1.2}
$$
\begin{array}{lcll}
\hA_1(\v{k}) &=& k \he_2(\v{k}) + \beta \hK_1(\v{k})\,, & \\
\hA_2(\v{k}) &=& k \he_1(\v{k}) + \beta \hK_2(\v{k})\,, & \quad \raisebox{0.3cm}[0cm]{$\v{k}\neq 0\,,$} \\
\hA_i(0) &=& \beta \hK_i(0)\,. &
\end{array}
\renewcommand{\arraystretch}{1}
$$
In the Schr{\"o}dinger representation, $\hA_i$ acts as
\renewcommand{\arraystretch}{1.9}
$$
\begin{array}{lcll}
\hA_1(\v{k}) &=& k \he_2(\v{k}) - \ds\i\hbar\frac{\kappa\beta}{2}\pdiff{}{e^*_1(\v{k})}\,, & \\
\hA_2(\v{k}) &=& k \he_1(\v{k}) - \ds\i\hbar\frac{\kappa\beta}{2}\pdiff{}{e^*_2(\v{k})}\,, & \quad \raisebox{0.4cm}[0cm]{$\v{k}\neq 0\,,$} \\
\hA_i(0) &=& \ds - \i\hbar\frac{\kappa\beta}{2}\pdiff{}{e^*_i(0)}\,. &
\end{array}
\renewcommand{\arraystretch}{1}
$$ 
Up to an $A$-dependent phase (which we choose to be zero), eigenstates of $\hA$ have the form
\be
\varphi_A[e_i(\v{k})] = \N\exp\left[\frac{2\i}{\hbar\kappa\beta}\sum_{\v{k}}\,\left(A^*_i(\v{k})e_i(\v{k}) - k\, e^*_1(\v{k})e_2(\v{k})\right)\right]\,.
\label{eigenstateA}
\ee
Within the quantum theory, the canonical transformation \eq{linearizedcanonicaltransformationclassical} is implemented by a unitary map
\bean
\psi[e] &\to& e^{\i f[e]/\hbar}\psi[e]\,, \\
\hat{O} &\to& e^{\i f[e]/\hbar}\,\hat{O}\,e^{-\i f[e]/\hbar}\,,
\eean
that turns the $\hA_i$-operator into a pure functional derivative in $e^*_i$, i.e.\
\be
\hA_i(\v{k}) \to -\i\hbar\frac{\kappa\beta}{2}\pdiff{}{e^*_i(\v{k})}\,.
\ee
We see from \eq{eigenstateA} that the required factor is
$$e^{\i f[e]/\hbar} := \exp\left(\frac{2\i}{\hbar\kappa\beta}\sum_{\v{k}}\,k\,e^*_1(\v{k})e_2(\v{k})\right)\,.$$
The transformed vacuum state reads
\be
\Psi[e_i(\v{k})] = \N\exp\left[-\frac{1}{\hbar\kappa}\sum_{\v{k}}\left(\omega(\v{k})\,e^*_i(\v{k})e_i(\v{k})
- \frac{2\i}{\beta}\,k\,e^*_1(\v{k})e_2(\v{k})\right)\right]\,,
\label{transformedgroundstate}
\ee
In terms of reduced Fourier components, it takes the form
\be
\Psi[e^{\sss\rm red}_{ab}(\v{k})] = \N\exp\left[-\frac{1}{\hbar\kappa}\sum_{\v{k}}\left(\omega(\v{k})\,e^{\sss\rm red*}_{ab}(\v{k})e^{\sss\rm red}_{ab}(\v{k})
+ \frac{1}{\beta}\,e^{\sss\rm red*}_{ab}(\v{k})\,\varepsilon_{acd} k_c e^{\sss\rm red}_{db}(\v{k})\right)\right]\,.
\label{groundstateFourier}
\ee
By doing a Gaussian integration, we can transform \eq{transformedgroundstate} to the $A$-representation:
\be
\Psi[A_i(\v{k})] = \N\exp\left\{-\frac{1}{\hbar\kappa}\left[\frac{1}{\beta^2\omega_0}\,A^2_i(0)
+ \frac{3/4}{1+\beta^2}\sum_{k>0}\,\frac{1}{k}
\left(A^*_i(\v{k})A_i(\v{k}) + \frac{2\i}{\beta}\,A^*_1(\v{k})A_2(\v{k})\right)\right]\right\}\,.
\label{groundstatepolarizationA}
\ee
The Schr{\"o}dinger representation we have defined so far is formal, since arbitrarily high momenta appear. 
We can resolve this either by a rigorous continuum formulation (employing Gaussian measures on tempered distributions), or by using a regularization. Here, we choose the second possibility, as the existence of infinitely high momenta is physically questionable anyhow. We introduce an ultraviolet cutoff on the momenta, which we denote by $\Lambda$. We will sometimes refer to this as the regularized Fock representation, since it is the natural home for Fock-like excitations, as opposed to the polymer-like excitations in the loop representation. (Of course, strictly speaking we are dealing with a Schr{\"o}dinger and not a Fock representation.)

With this adjustment, the canonically transformed and regularized vacuum becomes
\be
\Psi[e^{\sss\rm red}_{ab}(\v{k})] = \N\exp\left[-\frac{1}{\hbar\kappa}\sum_{k\le\Lambda}\left(\omega(\v{k})\,e^{\sss\rm red*}_{ab}(\v{k})e^{\sss\rm red}_{ab}(\v{k})
+ \frac{1}{\beta}\,e^{\sss\rm red*}_{ab}(\v{k})\,\varepsilon_{acd} k_c e^{\sss\rm red}_{db}(\v{k})\right)\right]\,.
\label{groundstateFourierregularized}
\ee
In the regularized scheme, we consider the position space field $e^{\sss\rm red}_{ab}(\v{x})$ as a function 
\be
e^{\sss\rm red}_{ab}(\v{x}) := \frac{1}{\sqrt{V}}\sum_{k\le\Lambda}\e^{\i\v{k}\cdot\v{x}}\,e^{\sss\rm red}_{ab}(\v{k})\,.
\ee
of the Fourier modes $e^{\sss\rm red}_{ab}(\v{k})$. With that in mind, we can write the state functional also as
\be
\Psi[e^{\sss\rm red}_{ab}(\v{k})] =
\N\exp\left[-\frac{1}{\hbar\kappa}\left(\int\d^3 x\int\d^3 y\;W_{\Lambda}(\v{x},\v{y})\,e^{\sss\rm red}_{ab}(\v{x})e^{\sss\rm red}_{ab}(\v{y})
- \frac{\i}{\beta}\int\d^3 x\;e^{\sss\rm red}_{ab}(\v{x})\,\varepsilon_{acd}\pa_c e^{\sss\rm red}_{db}(\v{x})\right)\right]\,,
\label{groundstateregularizedposition}
\ee
where the kernel $W_{\Lambda}$ is defined by
\be
W_{\Lambda}(\v{x},\v{y}) = \frac{1}{V}\sum_{k\le\Lambda}\e^{\i\v{k}\cdot(\v{x}-\v{y})}\omega(\v{k})\,.
\ee

\section{Transition to degrees of freedom of LQG}
\label{Transition_to_degrees_of_freedom_of_LQG}

At this point, we have a well-defined expression for a vacuum functional and we would like to translate it into a state of the LQG Hilbert space. In doing so, we want to preserve the physical properties of the Fock state as far as possible, which, includes, in particular, the cutoff on the momenta. The problem is that our state functional involves degrees of freedom that appear quite different from those of loop quantum gravity: on the one side, we have fields or their Fourier transforms, and on the other side abstract networks with spin labellings. In order to achieve a meaningful transition from Fock to loop state, it will be essential to find the right way to relate these degrees of freedom.

The Fock state was obtained by a reduced phase space quantization: we imposed the linearized gauge- and diff-constraint on the classical level, and then quantized the reduced degrees of freedom. Thus, we arrived at a state which is a functional of the reduced connection (see eqn.\ \eq{groundstatepolarizationA}).

LQG, on the other hand, is based on a quantization of the full phase space variables, yielding the kinematic Hilbert space $\H$. The full non-linear gauge- and diff-constraint are imposed subsequently to give the gauge- and diff-invariant Hilbert space $\H_0$ and $\H_{\rm diff}$ respectively. The configuration space, from which the Schr{\"o}dinger representation on $\H$ is built, consists of generalized connections --- distributional connections with support on graphs, of which ordinary connections are only a special case. 
States in $\H_{\rm diff}$ are functionals of gauge- and diff-equivalence classes of such generalized connections\footnote{The standard procedure is to define states in $\H_0$ as gauge-invariant functionals of generalized connections, and states in $\H_{\rm diff}$ are constructed as equivalence classes of states in $\H_0$ under diffeomorphisms. We obtain a mathematically equivalent formulation by defining everything in terms of equivalence classes of {\it connections}. Then, an element of $\H_{\rm diff}$ is a functional of equivalence classes of connections w.r.t.\ gauge- and diff- transformations.}.

Logically, we can divide this difference between degrees of freedom into three steps:
\begin{enumerate}
\item reduced connection $\to$ connection,
\item connection $\to$ generalized connection,
\item generalized connection $\to$ gauge- and diff-equivalence class of generalized connections. 
\end{enumerate}

Our strategy for bridging this gap:  
We modify the Fock state $\Psi$ such that it becomes a functional of connections (sec.\ \ref{From_reduced_to_full_configuration_space} and \ref{Complexifier_form}); then we switch from a pure momentum regularization to a combined momentum / triangulation based regularization which gives us $\Psi$ as a functional of generalized connections (sec.\ \ref{From_momentum_cutoff_to_triangulation}). In the final step, an averaging over the gauge- and diffeomorphism group has to be applied in order to arrive at a state in $\H_{\rm diff}$. In this paper, we only do the gauge-averaging explicitly, which provides a state in $\H_0$ (sec.\ \ref{Gauge_projection}). The diff-projection remains to be done.

\subsection{From reduced to full configuration space}
\label{From_reduced_to_full_configuration_space}

We take the Fourier coefficients $E^a_i(\v{k})$ as coordinates for the full configuration space. The position space field 
\be
E^a_i(\v{x}) = \frac{1}{\sqrt{V}}\sum_{k\le\Lambda}\e^{\i\v{k}\cdot\v{x}}\,E^a_i(\v{k})
\ee 
should be understood as a function of the $E^a_i(\v{k})$. Recall that $e_{ia}$ without index $^{\sss\rm red}$ denotes the difference
\be
e_{ia}(\v{x}) \equiv e^a_i(\v{x}) = E^a_i(\v{x}) - \delta^a_i
\ee
between the background triad and $E^a_i$. 
We introduce a Schr{\"o}dinger representation for functionals on the full configuration space: the measure is defined
by
\be
\int DE := \prod_{k\le\Lambda,\: k^1\ge 0}\;\prod_{i=1}^3\prod_{a=1}^3\prod_{r=0}^1\int_{-\infty}^\infty\d E^a_{ir}(\v{k})\,,
\ee
and we represent the canonical commutation relations 
\be
\Big[\hE^a_i(\v{k}),\hA^{k\dagger}_b(\v{k}')\Big] = \i\hbar\frac{\beta\kappa}{2}\,\delta^b_a\delta^k_i\,\delta_{\v{k},\v{k}'}\,.
\ee
(cf.\ \eq{newPoissonbracketFourier}) by setting
\be
\hE^a_i(\v{k}) = E^a_i(\v{k})\quad\mbox{and}\quad \hA^i_a(\v{k}) = -\i\hbar\frac{\kappa\beta}{2}\,\pdiff{}{E^{a*}_i(\v{k})}\,.
\ee
Analogously, there is an $A$-representation, where $\hE^a_i(\v{k})$ and $\hA^i_a(\v{k})$ are represented by
\be
\hA^i_a(\v{k}) = A^i_a(\v{k})\quad\mbox{and}\quad \hE^a_i(\v{k}) =  \i\hbar\frac{\kappa\beta}{2}\,\pdiff{}{A^{i*}_a(\v{k})}\,.
\label{Arepresentation}
\ee
To start with, we simplify the state \eq{groundstateregularizedposition} by dropping the $\beta$-dependent phase factor:
\be
\Psi[e^{\sss\rm red}_{ab}(\v{k})] =
\N\exp\left[-\frac{1}{\hbar\kappa}\int\d^3 x\int\d^3 y\;W_\Lambda(\v{x},\v{y})\,e^{\sss\rm red}_{ab}(\v{x})e^{\sss\rm red}_{ab}(\v{y})\right]\,.
\label{betatermdropped}
\ee
Thus, we avoid overly long formulas in the computations that follow.
The treatment {\it with} phase factor is discussed in section \ref{inclusionofphasefactor}.

Our aim is to extend the functional \eq{betatermdropped} to the full configuration space.
The most simple possibility would be to use the projection map
\be
e^{\sss\rm red}_{ab}(\v{k}) = P_{ab}^{cd}(\v{k})e_{cd}(\v{k})\,, 
\ee
and define the extended state by the pull-back, i.e.
\be
\Psi_{\rm ext}[E^a_i(\v{k})] :=
\N\exp\left[-\frac{1}{\hbar\kappa}\int\d^3 x\int\d^3 y\;W_\Lambda(\v{x},\v{y})\,(Pe)_{ab}(\v{x})(Pe)_{ab}(\v{y})\right]\,.
\label{keepingprojectors}
\ee  
The problem with this state is that it has a very degenerate peak. In the vicinity of the background triad, this is ok because it corresponds to gauge-, diff- and time reparametrization invariance. If one goes farther away from the background triad, however, the linearized transformations are no longer symmetries of the theory. That means that if we follow long enough along the degenerate direction, we will arrive at triads that are very diff- and gauge-{\it in}equivalent to the chosen background, but they are still in the peak of the state functional due to the projector. That is a very unphysical property.

As an alternative, we could drop the projectors in \eq{keepingprojectors} and define the state as
\be
\Psi_{\rm ext}[E^a_i(\v{k})] :=
\N\exp\left[-\frac{1}{\hbar\kappa}\int\d^3 x\int\d^3 y\;W_\Lambda(\v{x},\v{y})\,e_{ab}(\v{x})e_{ab}(\v{y})\right]\,.
\label{droppingprojectors}
\ee  
It implies that we make the Gaussian peak non-degenerate and throw out diff- and gauge-symmetry completely. 
After the state has been transferred to the LQG Hilbert space, the lost symmetries need to be restored. We can do this by applying an averaging over the gauge and diffeomorphism group\footnote{By dropping the projector in \eq{droppingprojectors}, we also lost the linearized scalar constraint: it is related to the dynamics and should reappear when the dynamics is linearized at the level of loops. For ideas in this direction, see section \ref{freetheoryperturbationtheoryandrenormalization}.}. If we did the averaging in \eq{droppingprojectors}, it would effectively reintroduce the projection on the symmetric transverse part for small $e$, while for large $e$, it would establish the correct non-linear gauge- and diff-symmetry for the state. The problem: at the level of the loop state, gauge-averaging turns out to be rather complicated, if we start from \eq{droppingprojectors}. 

In the present paper, we use a variation of this approach: to simplify the gauge-averaging, we replace the 
triad fluctuations $e_{ab}(\v{x})$ in \eq{droppingprojectors} by the fluctuation of the densitized inverse metric\footnote{We thank C.\ Rovelli for suggesting this modification.} $\tilde{g}^{ab}(\v{x})$: 
\be 
e_{ab}(\v{x})\quad \to \quad\frac{1}{2}\left(\tilde{g}^{ab}(\v{x}) - \delta^{ab}\right) = \frac{1}{2}\left(E^a_i(\v{x})E^b_i(\v{x}) - \delta^{ab}\right)\,.
\label{replacementestt}
\ee
The triad fields are 1-densitites, so $\tilde{g}^{ab}(\v{x})$ has density weight 2.
Since $E^a_i$ contains only modes up to $k=\Lambda$, we can write $\tilde{g}^{ab}(\v{x})$ also in a smeared form 
\be
\tilde{g}^{ab}_{\Lambda} := \int\d^3 x'\; E^a_i(\v{x}) \delta_\Lambda(\v{x}-\v{x}') E^b_i(\v{x}')\,,
\ee
where the smearing is done with the regularized delta function
\be
\delta_\Lambda(\v{x}) := \frac{1}{V}\sum_{k\le\Lambda}\e^{\i\v{k}\cdot\v{x}}\,.
\ee
The new state $\Psi_{\rm ext}$ is defined as
\bea
\Psi_{\rm ext}[E^a_i(\v{k})] &:=&
\N\exp\Bigg[-\frac{1}{4\hbar\kappa}\int\d^3 x\int\d^3 y\;W_\Lambda(\v{x},\v{y})\,
\left(\int\d^3 x'\;E^a_i(\v{x})\delta_\Lambda(\v{x}-\v{x}')E^b_i(\v{x}') - \delta^{ab}\right) \nonumber \\
&& \hspace{6cm}\times\,\left(\int\d^3 y'\;E^a_i(\v{y})\delta_\Lambda(\v{y}-\v{y}')E^b_i(\v{y}') - \delta^{ab}\right)\Bigg] 
\label{modifiedstate}\,.
\eea
Note that this state is almost gauge-invariant, but not completely, due to the smearing at the cutoff scale.

What is the justification for changing from the state \eq{droppingprojectors} to \eq{modifiedstate}? With our substitution \eq{replacementestt}, the term
\be
e_{ab}(\v{x})e_{ab}(\v{y})
\label{esquare}
\ee
becomes
\bea
\lefteqn{\frac{1}{2}\left(\int\d^3 x'\; E^a_i(\v{x}) \delta_{\Lambda}(\v{x}-\v{x}') E^b_i(\v{x}') - \delta^{ab}\right)
\frac{1}{2}\left(\int\d^3 y'\; E^a_k(\v{y}) \delta_{\Lambda}(\v{y}-\v{y}') E^b_k(\v{y}') - \delta^{ab}\right)} \label{replacementesquare} \\
&&= \frac{1}{4}\left(E^a_i(\v{x})E^b_i(\v{x}) - \delta^{ab}\right)
\left(E^a_k(\v{y})E^b_k(\v{y}) - \delta^{ab}\right) \nonumber \\
&&=
\frac{1}{4}\left[\Big(\delta^a_i + e^a_i(\v{x})\Big)\left(\delta^b_i + e^b_i(\v{x})\right) - \delta^{ab}\right]
\left[\left(\delta^a_k + e^a_k(\v{y})\right)\left(\delta^b_k + e^b_k(\v{y})\right) - \delta^{ab}\right] \nonumber \\
&&=
\frac{1}{4}\left[\delta^a_i\,e^b_i(\v{x})
+ e^a_i(\v{x})\,\delta^b_i
+ e^a_i(\v{x})\,e^b_i(\v{x})\right]\left[\delta^a_k\,e^b_k(\v{y})
+ e^a_k(\v{y})\,\delta^b_k
+ e^a_k(\v{y})\,e^b_k(\v{y})\right] \nonumber \\
&&= \frac{1}{4}\left(e^b_a(\v{x}) + e^a_b(\v{x})\right)
\left(e^b_a(\v{y}) + e^a_b(\v{y})\right) + o(e^3) \nonumber \\
&&= e_{(ab)}(\v{x})\,e_{(ab)}(\v{y}) + o(e^3)\,.
\label{plusthirdorder}
\eea
The basic premise of linearization is that for a suitable coupling parameter, the values of $e$ can be divided into ``large'' and ``small'' fluctuations with the following property: the ``large'' $e$ are irrelevant because for these values the state functionals under consideration are exponentially damped. The remaining ``small'' fluctuations lie near the peak of the states and are small {\it enough} that to first approximation, higher orders in $e$ can be neglected relative to the leading order terms. Thus, for small $e$, and within the precision of the linear approximation, \eq{replacementesquare} and \eq{esquare} are equal except for the symmetrizers --- a degeneracy which is due to the gauge-invariance in \eq{modifiedstate}. Since we intend to apply a gauge-averaging anyhow, we can ignore this difference. For large $e$, both state functionals are exponentially damped, so that again the difference between  \eq{replacementesquare} and \eq{esquare} is not important.

\subsection{``Complexifier'' form}
\label{Complexifier_form}

By a Gaussian integration, we transform the state \eq{modifiedstate} to the $A$-representation:
\bea
\Psi_{\rm ext}[A^i_a(\v{k})] &=& \N\,\int DE\;
\exp\left(-\frac{2\i}{\hbar\kappa\beta}\sum_{k\le\Lambda} A^{a*}_i(\v{k}) E^a_i(\v{k})\right) \nonumber \\
&& \times\,\exp\Bigg[-\frac{1}{4\hbar\kappa}\int\d^3 x\int\d^3 y\;W_\Lambda(\v{x},\v{y})\,
\left(\int\d^3 x'\;E^a_i(\v{x})\delta_\Lambda(\v{x}-\v{x}')E^b_i(\v{x}') - \delta^{ab}\right) \nonumber \\
&& \hspace{6cm}\times\,\left(\int\d^3 y'\;E^a_i(\v{y})\delta_\Lambda(\v{y}-\v{y}')E^b_i(\v{y}') - \delta^{ab}\right)\Bigg]\,.
\label{functionalofAbeforecomplexifier}
\eea
%\left(\prod_{k\le\Lambda,\: k^1>0}\;\prod_{i=1}^3\prod_{a=1}^3\prod_{r=0}^1\int_{-\infty}^\infty\d e^a_{ir}(\v{k})\right)
Using that
\bea
\lefteqn{\int DE\;
\exp\left(-\frac{2\i}{\hbar\kappa\beta}\sum_{k\le\Lambda} A^{a*}_i(\v{k}) E^a_{ir}(\v{k})\right)} \nonumber \\
&=& 
\left(\prod_{k\le\Lambda,\: k^1>0}\;\prod_{i=1}^3\prod_{a=1}^3\prod_{r=0}^1\int_{-\infty}^\infty\d E^a_{ir}(\v{k})\right)
\exp\left(-\frac{2\i}{\hbar\kappa\beta}\sum_{k\le\Lambda} A^a_{ir}(\v{k}) E^a_{ir}(\v{k})\right) \nonumber
\eea
is the delta functional on the connection, and that the operator $\hE^{a\dagger}_i$ acts like 
$$
\i\hbar\frac{\kappa\beta}{2}\,\pdiff{}{A^i_a(\v{k})}\,,
$$
we can write the entire expression \eq{functionalofAbeforecomplexifier} as 
\bea
\Psi_{\rm ext}[A^i_a(\v{k})] &=& \N\exp\Bigg[-\frac{1}{4\hbar\kappa}\int\d^3 x\int\d^3 y\;W_\Lambda(\v{x},\v{y})\,
\left(\int\d^3 x'\;\hE^a_i(\v{x})\delta_\Lambda(\v{x}-\v{x}')\hE^b_i(\v{x}') - \delta^{ab}\right) \nonumber \\
&& \hspace{6cm}\times\,\left(\int\d^3 y'\;\hE^a_i(\v{y})\delta_\Lambda(\v{y}-\v{y}')\hE^b_i(\v{y}') - \delta^{ab}\right)\Bigg]\;\delta(A)\,, \nonumber \\
&& \label{complexifierform}
\eea
where 
\be
\hE^a_i(\v{x}) = \frac{1}{\sqrt{V}}\sum_{k\le\Lambda}\e^{\i\v{k}\cdot\v{x}}\,\hE^a_i(\v{k})\,.
\ee 
This form of the state is similar, {\it but not identical}, to Thiemann's general complexifier form for coherent states \cite{Treviewcoherentstates}. Thiemann writes coherent states as
\be
\exp\left(-\hat{C}\right)\,\delta(A-\tilde{A})_{\Big|_{\tilde{A}\to A^{\bC}_{cl}}}\,
\ee
where $A^{\bC}_{cl}$ is the so-called complexified connection and contains the background triad.
What we do here is somewhat different because we leave the background triad {\it outside} the delta-function. 
For that reason, it is a slight abuse of terminology if we call our analogue of $\hat{C}$, i.e.
\bea
\hat{C} &:=& 
\frac{1}{4\hbar\kappa}\int\d^3 x\int\d^3 y\;W_\Lambda(\v{x},\v{y})\,
\left(\int\d^3 x'\;\hE^a_i(\v{x})\delta_\Lambda(\v{x}-\v{x}')\hE^b_i(\v{x}') - \delta^{ab}\right) \nonumber \\
&& \hspace{4.2cm}\times\,\left(\int\d^3 y'\;\hE^a_i(\v{y})\delta_\Lambda(\v{y}-\v{y}')\hE^b_i(\v{y}') - \delta^{ab}\right)\,,
\eea
again a complexifier. 

\subsection{From momentum cutoff to triangulation}
\label{From_momentum_cutoff_to_triangulation}

At this point, we have a well-defined expression for a state functional of connection Fourier modes. In the next step, we turn this into a functional of generalized connections. We do so by a change of regularization scheme: similarly as one changes from a momentum to a lattice regularization in ordinary QFT, we trade the UV cutoff on the connection for a triangulation of space. Let $\T_{\Lambda}$ denote the simplicial complex of this triangulation. The connection is replaced by a generalized connection $\Ab$ on the dual complex\footnote{Given $\T_{\Lambda}$, the dual complex is defined in the standard way, using the metric information of the background.} $\T^*_{\Lambda}$: a map that sends every edge $e$ of $\T^*_{\Lambda}$ into a group element $g_e = \Ab(e)$. In making this transition, we want to alter the physical properties of the state as little as possible. The triangulation-based regularization should be such that it mimics the effects of the UV cutoff. For that reason, we choose $\T_{\Lambda}$ to be regular in the following sense: 

When measured against the background metric $\delta_{ab}$, the edges $e$ of the dual complex should be straight and have lengths $l_e$ in the range 
$$(1-\epsilon) l_{\Lambda}\;\;\le\;\; l_e\;\; \le\;\; (1+\epsilon)l_{\Lambda}\,,$$
where 
\be
l_{\Lambda} := \frac{\pi}{\Lambda}
\ee
is the length scale corresponding to the cutoff $\Lambda$, and $\epsilon$ is some small fixed number. 
\label{definitionofdualcomplex}

The state we have so far consists of two parts: a delta functional of the Fourier coefficients of the connection and an operator acting on it. Let us first consider the delta functional: we replace it by the delta functional $\delta_{\T^*}(\Ab)$ on $\T^*_{\Lambda}$, which is equal to a sum over all gauge-{\it variant} spin networks $\St$ restricted to $\T^*_{\Lambda}$:
\be
\delta_{\T^*}\left(\Ab\right)
= \sum_{\St\subset\T^*_{\Lambda}} \St(0)\St^*(\Ab)
\label{deltafunctional}
\ee
In the operator part in \eq{complexifierform}, we have to replace the smeared operator product 
by an operator on functionals of $\Ab$, or equivalently, by an operator on spin network states. 
In other words: after having quantized 
\be
\tilde{g}^{ab}_{\Lambda} = \int\d^3 x'\;E^a_i(\v{x})\delta_\Lambda(\v{x}-\v{x}')E^b_i(\v{x}')
\label{smearedmetric}
\ee
in a Fock space manner, we will now quantize it along the lines of loop quantum gravity.

\subsubsection*{Loop quantization of smeared inverse densitized metric}

There are several consistent ways in which one could quantize expression \eq{smearedmetric} on a triangulation. Of all the possibilities we will choose one that is very simple and gauge-symmetric. 

Before dealing directly with \eq{smearedmetric}, let us give a definition for $E^a_i(\v{x})$. Spin network states are built from representation matrices $U_j(g_e)$ associated to edges $e$. To define $\hE^a_i(\v{x})$ on $U_j(g_e)$, we think of the latter as the path-ordered exponential of a connection
$A^i_a(\v{x})$, i.e.
\bean
\lefteqn{U_j(g_e) =
\P\exp\left(\frac{\i}{\hbar}\int_{e}\d s\;\dot{e}^a(s)A^i_a(\v{e}(s))\stackrel{(j)}{J_i}\right)} \nonumber \\
&&= \sum_{n=0}^\infty\,\left(\frac{\i}{\hbar}\right)^n\int\limits_0^1\d s_1\int\limits_0^{s_1}\d s_2\cdots\int\limits_0^{s_{n-1}}\!\!\d s_n\;
\dot{e}^{a_1}(s_1)A^{i_1}_{a_1}(\v{e}(s_1))\cdots\dot{e}^{a_n}(s_n)A^{i_n}_{a_n}(\v{e}(s_n))
\stackrel{(j)}{J_{i_1}}\cdots\stackrel{(j)}{J_{i_n}}\,,
\eean
and view the triad operator as the functional derivative $\i\hbar\kappa\beta/2\,\delta/\delta A^i_a(\v{x})\,.$ With this prescription, we obtain that
\be
\label{actionEonU}
\hE^a_i(\v{x})U_j(g_e) = -\frac{\kappa\beta}{2}\,\int_0^1\d s\;\dot{e}^a(s)\,\delta(\v{x}-\v{e}(s))\;U_j(g_{e_1(\v{x})})\!\stackrel{(j)}{J_i}U_j(g_{e_2(\v{x})})\,,
\ee
where $e_1(\v{x})$ and $e_2(\v{x})$ are the edges which result from splitting the edge $e$ at the point $\v{x}$. (When $\v{x}$ is not on $e$, the definition of $e_1(\v{x})$ and $e_2(\v{x})$ is irrelevant, since the delta function gives zero.) The problem with \eq{actionEonU} is that the original holonomy is split into two holonomies. Thus, when applied to a spin network on $\T^*_{\Lambda}$, the result will be a spin network which lives on the dual of a refined triangulation. If we want to stick to our original intention of defining the state on $\T^*_{\Lambda}$, we have to modify the action \eq{actionEonU} such that it leaves the space of spin networks on $\T^*_{\Lambda}$ invariant.

\psfrag{e}{$\sss e$}
\psfrag{v}{$\sss v$}
\psfrag{J}{$\sss J_i$}
\psfrag{J2}{$\sss J_i$}
\psfrag{e1}{$\sss e_1$}
\psfrag{e2}{$\sss e_2$}
The choice we take is
\be
\hE^a_i(\v{x})U_j(g_e) = -\frac{\kappa\beta}{4}\,\left(\int_0^1\d s\;\dot{e}^a(s)\right) \left(\delta(\v{x}-\v{e}(1))\stackrel{(j)}{J_i}U_j(g_{e}) 
+ \delta(\v{x}-\v{e}(0))\;U_j(g_{e})\!\stackrel{(j)}{J_i}\right)\,.
\label{actionEonUmodified}
\ee
This means that $\hE^a_i$-operators can only create $J$'s at the beginnings and ends of edges.
In the obvious way, equation \eq{actionEonUmodified} generalizes to an action of $\hE^a_i(\v{x})$ on an entire spin network state on $\T^*_{\Lambda}$:
\be
\hE^a_i(\v{x})\;\St = 
\sum_v\!\!\!\sum_{\parbox{1.3cm}{\centering {\tiny edges} \\ \vspace{-2mm} $\ss e$ {\tiny of} $\ss v$}} \!\!\! F^a_{v,e}(\v{x})\;\parbox{2cm}{\includegraphics[height=2cm]{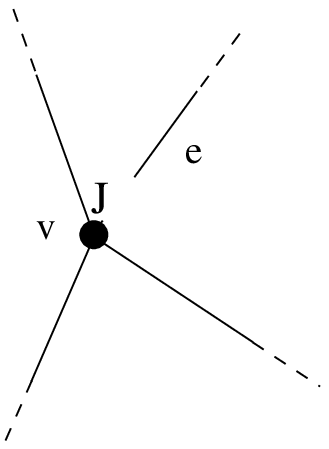}}\,.
\ee
The sum ranges over all vertices of the spin network, and the diagrams indicate where the $J$'s are inserted.
The {\it vertex-edge} form factor $F^a_{v,e}$ stands for 
\be
F^a_{v,e}(\v{x}) := -\frac{\kappa\beta}{4}\left(\int_0^1\d s\;\dot{e}^a(s)\right)\delta(\v{x}-\v{x}_v)\,.
\label{vertexedgeformfactor}
\ee
 
Given this definition of $\hE^a_i(\v{x})$, what would be a meaningful way to implement
\be
\hat{\tilde{g}}^{ab}_{\Lambda} \equiv 
\int\d^3 x'\;\hE^a_i(\v{x})\delta_{\Lambda}(\v{x}-\v{x}')\hE^b_i(\v{x}')\quad ?
\label{smearedmetricoperator}
\ee
Roughly speaking, the smearing function $\delta_{\Lambda}(\v{x}-\v{x}')$ requires $\v{x}$ and $\v{x}'$ to be closer than $l_{\Lambda}$.
Since the cutoff length $l_{\Lambda}$ is also the length scale of the dual edges, we translate this into the condition that, if $\hE^b_i(\v{x}')$ inserts a $J$ at a node $v$, $\hE^a_i(\v{x})$ can only insert $J$'s at the same node.

This still allows for the possibility that $\hE^a_i(\v{x})$ and $\hE^b_i(\v{x}')$ insert $J$'s on different edges of the same node. The action of the operator \eq{smearedmetricoperator} becomes
\bea
\lefteqn{\int\d^3 x'\;\hE^a_i(\v{x}) \delta_{\Lambda}(\v{x}-\v{x}') \hE^b_i(\v{x}')\;\St} \nonumber \\
&&= \sum_v\!\!\!\sum_{\parbox{1.3cm}{\centering {\tiny edges} \\ \vspace{-2mm} $\ss e_1,e_2$ {\tiny of} $\ss v$}} 
\!\!\!\int\d^3 x'\; F^a_{v,e_1}(\v{x})\delta_{\Lambda}(\v{x}-\v{x}')F^a_{v,e_2}(\v{x}')
\;\parbox{2cm}{\includegraphics[height=2cm]{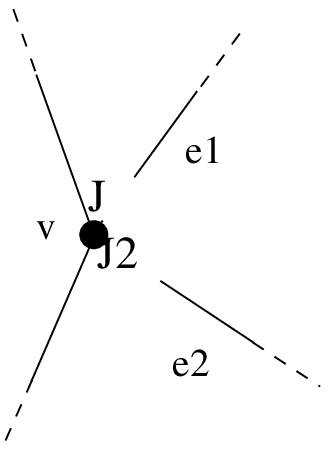}}
\,.
\label{smearedmetricoperatoronStwithcrossterms}
\eea
Due to the contraction of $i$-indices, this action is gauge-invariant. For a generic basis of spin networks, it is not diagonal, however, and that makes its use complicated.
In the following, we take a simpler choice, and adopt the point of view that $\hE^a_i(\v{x})\hE^b_i(\v{x}')$ should {\it not} involve contractions of $J_i$'s from different edges. If we drop these cross-terms, the action of \eq{smearedmetricoperator} is diagonal and simply reads
\be
\int\d^3 x'\;\hE^a_i(\v{x}) \delta_{\Lambda}(\v{x}-\v{x}') \hE^b_i(\v{x}')\;\St 
= \gt^{ab}_{\tilde{S}}(\v{x})\,\St\,,
\label{smearedmetricoperatoronSt}
\ee
where $\gt^{ab}_{\tilde{S}}(\v{x})$ is given by
\bea
\gt^{ab}_{\tilde{S}}(\v{x}) &:=& \sum_v\sum_{\mbox{$\ss e$ {\tiny of} $\ss v$}}
\int\d^3 x'\;F^a_{v,e}(\v{x})\delta_{\Lambda}(\v{x}-\v{x}')F^b_{v,e}(\v{x}')
\,\hbar^2 j_e(j_e+1) \nonumber \\
&\equiv& \sum_v\sum_{\mbox{$\ss e$ {\tiny of} $\ss v$}} F^{ab}_{v,e}(\v{x})\, j_e(j_e+1)\,,
\label{metricassociatedtoSt}
\eea
and 
\bea
F^{ab}_{v,e}(\v{x}) &:=& \int\d^3 x'\; \hbar^2 F^a_{v,e}(\v{x})\delta_\Lambda(\v{x}-\v{x}') F^b_{v,e}(\v{x}') \nonumber \\
&=&  
\left(\frac{\hbar\kappa\beta}{4}\right)^2
\left(\int_0^1\d s\;\dot{e}^a(s)\right)\left(\int_0^1\d s'\;\dot{e}^b(s')\right)
\int\d^3 x'\;\delta(\v{x}-\v{x}_v)\,\delta_{\Lambda}(\v{x}-\v{x}')\,\delta(\v{x}'-\v{x}_v) \nonumber \\
&=& 
\left(\frac{l_p^2\beta}{4}\right)^2
\int_0^1\d s\int_0^1\d s'\;\;\dot{e}^a(s)\dot{e}^b(s')\delta_{\Lambda}(0)\,\delta(\v{x}-\v{x}_v)\,.
\label{vertexformfactor}
\eea
We interpret $\gt^{ab}_{\St}$ as a densitized inverse metric that is associated to the spin network $\St$. This metric is distributional and has only support on vertices of the spin network graph. Thus, we can also write
\be
\gt^{ab}_{\tilde{S}}(\v{x}) = \sum_v\,\gt^{ab}_{\tilde{S}}(v)\,\delta(\v{x}-\v{x}_v)
\ee
where
\be
\gt^{ab}_{\tilde{S}}(v) := \frac{\beta^2}{16}\,l_p^4\delta_\Lambda(0)\;\sum_{\mbox{$\ss e$ {\tiny of} $\ss v$}}j_e(j_e+1)\int_0^1\d s\int_0^1\d s'\;\;\dot{e}^a(s)\dot{e}^b(s')\,.
\ee

With the definition \eq{smearedmetricoperatoronSt} and the replacement of the delta functional by \eq{deltafunctional}, we obtain the state
\bea
\Psi &=& \N \sum_{\St\subset\T^*_{\Lambda}} \St(0)
\exp\Bigg[-\frac{1}{4\hbar\kappa}\int\d^3 x\int\d^3 y\;W_\Lambda(\v{x}-\v{y})
\left(\gt^{ab}_{\tilde{S}}(\v{x}) - \delta^{ab}\right) 
\left(\gt^{ab}_{\tilde{S}}(\v{y}) - \delta^{ab}\right)
\Bigg]\;\St^* \\
&=& \N \sum_{\St\subset\T^*_{\Lambda}} \St(0)
\exp\Bigg[-\frac{1}{4\hbar\kappa}\int\d^3 x\int\d^3 y\;W_\Lambda(\v{x}-\v{y})
\,\htil^{ab}_{\tilde{S}}(\v{x})\,\htil^{ab}_{\tilde{S}}(\v{y})
\Bigg]\;\St^*\,.
\label{diagonalcontributiondropped}
\eea
In the second line, we abbreviated
\be
\htil^{ab}_{\St} := \gt^{ab}_{\St} - \delta^{ab}\,.
\label{fluctuationhS}
\ee
Since spin networks on $\T^*_{\Lambda}$ are naturally mapped to spin networks in $\H$, the state \eq{diagonalcontributiondropped} trivially extends to a state in $\H$. We denote this extended state by the same formula. Note that the restriction to spin networks on $\T^*_{\Lambda}$ in $\H$ is not ad hoc, but designed to preserve the momentum cutoff of the original Fock state. At the same time, the momentum regularization is not replaced completely, as it is still present in the kernel $W_\Lambda$ and the factor $\delta_{\Lambda}(0)$.

\subsection{Gauge projection}
\label{Gauge_projection}

It remains to make the transition from \eq{diagonalcontributiondropped} to a gauge-invariant state in $\H_0$.
Gauge-averaging simply yields % Argument mit delta-Funktion auf Raum der Funktionen mit fixiertem Spin-labelling.
\be
\Psi_0 = \N \sum_{S\subset\T^*_{\Lambda}} S(0)
\exp\Bigg[-\frac{1}{4\hbar\kappa}\int\d^3 x\int\d^3 y\;W_\Lambda(\v{x}-\v{y})
\,\htil^{ab}_S(\v{x})\,\htil^{ab}_S(\v{y})\Bigg]\; S^*\,,
\label{gaugeinvariantstate} 
\ee
where the sum ranges over all gauge-{\it invariant} spin networks $S$ on $\T^*_{\Lambda}$. We introduce the coefficient
\be
\Psi_0(S) := S(0)
\exp\Bigg[-\frac{1}{4\hbar\kappa}\int\d^3 x\int\d^3 y\;W_\Lambda(\v{x}-\v{y})
\,\htil^{ab}_S(\v{x})\,\htil^{ab}_S(\v{y})\Bigg]\,,
\label{spinnetworkcoefficient}
\ee
and write this more compactly as 
\be
\Psi_0 = \N \sum_{S\subset\T^*_{\Lambda}}\Psi_0(S)S^*\,. 
\ee
One can think of $\Psi_0(S)$ as the wavefuntion of $\Psi_0$ in the $S$-representation.

When expressed in terms of Fourier coefficients of the two-tangent form factors $F^{ab}_{v,e}$, the coefficient reads
\be
\Psi_0(S) = S(0)
\exp\left[-\frac{1}{4 l_p^2}
\sum_{k\le\Lambda}\,\omega(\v{k})\left|\sum_v\sum_{\mbox{$\ss e$ {\tiny of} $\ss v$}} F^{ab}_{v,e}(\v{k})\,j_e(j_e+1) - \sqrt{V}\delta^{ab}\delta_{\v{k},0}\right|^2\right]\,.
\label{spinnetworkcoefficientFourier}
\ee
For $\Psi_0$ to be a well-defined state in the gauge-invariant Hilbert space $\H_0$, the norm
\be
\|\Psi_0\|^2 = \N^2 \sum_{S\subset\T^*_{\Lambda}} |S(0)|^2
\exp\left[-\frac{1}{2 l_p^2}
\sum_{k\le\Lambda}\,\omega(\v{k})\left|\sum_v\sum_{\mbox{$\ss e$ {\tiny of} $\ss v$}} F^{ab}_{v,e}(\v{k})\,j_e(j_e+1) - \sqrt{V}\delta^{ab}\delta_{\v{k},0}\right|^2\right]
\label{normofstate}
\ee
has to be finite.

The number of possible spin network graphs on $\T^*_{\Lambda}$ is finite. Therefore, in order to show that 
\eq{normofstate} is finite, it suffices to prove that for every graph $\gamma$ on $\T^*_{\Lambda}$,
\be
\sum_{\parbox{1.2cm}{\centering \tiny $S$ with \\ graph $\gamma$}}
|S(0)|^2 \exp\left[-\frac{1}{2 l_p^2}
\sum_{k\le\Lambda}\,\omega(\v{k})\left|\sum_v\sum_{\mbox{$\ss e$ {\tiny of} $\ss v$}}F^{ab}_{v,e}(\v{k})\,j_e(j_e+1) - \sqrt{V}\delta^{ab}\delta_{\v{k},0}\right|^2\right] < \infty\,.
\label{fixedgraph}
\ee

Let us abbreviate $j_e(j_e+1)$ by $c_e$. The factor $S(0)$ gives a polynomial in the $j_e$'s that depends on the connectivity of the graph $\gamma$. For a suitable polynomial $P[c_e]$ in $c_e$, we have
\be
|S(0)|^2 \le P[c_e]\,.
\ee 
Thus, an upper bound on \eq{fixedgraph} is given by
\be
\left(\prod_e\sum_{j_e = 1/2}^\infty\right) P[c_e] \exp\left(-\sum_{ee'}M_{ee'}\,c_e c_{e'} + \sum_e N_e c_e + K\right)\,,
\ee
where $M_{ee'}$ is a symmetric positive matrix. By a linear transformation 
\be
\tilde{c}_e = T_{ee'}c_{e'}\,,
\ee
we render $M_{ee'}$ diagonal and arrive at the upper estimate
\be
\left(\prod_e\sum_{\tilde{c}_e = 0}^\infty\right) \tilde{P}[\tilde{c}_e] \exp\left(-\sum_e \lambda_e \tilde{c}^2_e + \sum_e \tilde{N}_e \tilde{c}_e\right)\,,
\label{upperestimate}
\ee
where the eigenvalues $\lambda_e$ are positive and $\tilde{P}$ is a polynomial in $\tilde{c}_e$. By using further estimates with integrals, \eq{upperestimate} can be shown to be convergent.

We conclude that the state \eq{gaugeinvariantstate} is an element in the gauge-invariant Hilbert space $\H_0$.

% Anmerkungen zum Beweis:

% - Warum Imagin{\"a}r-Teil in \tilde{N}_e kein Problem ist:
% |int_{\bR} exp(-x^4 + ix^2)| \le int_{\bR} exp(-x^4)
% - Warum Gemischtheit der Polynome kein Problem ist:
% x^k y^l \le x^n + y^m, wenn n und m gro{\ss} genug gew{\"a}hlt sind.
% - {\"U}bergang von diskreter Summe zu Integralen:
% diskrete Summe 
% sum_{n=0}^infty n^4 exp(-n^2)
% als Integration {\"u}ber Treppenfunktion betrachten und geeignet
% mit kontinuierlicher Funktion absch{\"a}tzen; f{\"u}hrt auf eine Funktion
% der Form x^4 exp(-x^2) oder so {\"a}hnlich, evt.
% mit Translation / Faktoren.

\subsection{Inclusion of phase factor}
\label{inclusionofphasefactor}

Equation \eq{gaugeinvariantstate} is not yet the final form of the LQG state: what is missing is the contribution from the phase factor
\be
\exp\left[\frac{\i}{\hbar\kappa\beta}\int\d^3 x\;e^{\sss\rm red}_{ab}(\v{x})\,\varepsilon_{acd}\pa_c e^{\sss\rm red}_{db}(\v{x})\right]
\ee
that we have ignored so far in our transition from Fock to loop state. The treatment of the additional integral
is quite analogous to that of the Gaussian term. There is only one difference --- a difficulty that arises when implementing the differential operator $\pa_c$ on the triangulation $\T^*_\Lambda$. If we were using a hypercubic lattice, we would simply get
\bea
\Psi_0(S) &=& S(0)
\exp\Bigg[-\frac{1}{4\hbar\kappa}\int\d^3 x\int\d^3 y\;W_\Lambda(\v{x}-\v{y})
\,\htil^{ab}_S(\v{x})\,\htil^{ab}_S(\v{y}) \nonumber \\
&& \hspace{1.7cm}{}+
\frac{\i}{4\hbar\kappa\beta}\sum_v 
\,\htil^{ab}_S(v)
\,\varepsilon^{acd}\,\nabla^c
\htil^{ab}_S(v)
\Bigg]\,,
\eea
where $\nabla^c$ stands for the lattice derivative in the $c$-direction. On a triangulation an implementation of $\pa_c$ is more complicated and requires additional weighting factors to take account of the geometry of the triangulation. We do not determine this in detail and content ourselves with saying that the final state has the wavefunction
\bea
\Psi_0(S) = S(0)
\exp\Bigg[-\frac{1}{4\hbar\kappa}\int\d^3 x\int\d^3 y\;W_\Lambda(\v{x}-\v{y})
\,\htil^{ab}_S(\v{x})\,\htil^{ab}_S(\v{y}) \quad + \quad\mbox{phase term}\quad\Bigg]
\label{spinnetworkcoefficientwithphasefactor}
\eea
where the phase term is an analogue of 
\be
\frac{\i}{4\hbar\kappa\beta}\sum_v 
\,\htil^{ab}_S(v)
\,\varepsilon^{acd}\,\nabla^c
\htil^{ab}_S(v)
\ee 
on a triangular lattice.

\section{Graviton states}
\label{Graviton_states}

The steps that led us from the vacuum of linearized gravity to the state $\Psi_0$ can be repeated in complete analogy for gravitons. Let us go back to the Schr{\"o}dinger representation of linearized extended ADM gravity (see \ref{Reduced_phase_space_quantization}). For $\v{k} \neq 0$, we define creation and annihilation operators
\bea
a_i(\v{k}) &:=& \sqrt{\frac{k}{\hbar\kappa}}\,e_i(\v{k}) + \frac{\i}{\sqrt{\hbar\kappa k}}\,K_i(\v{k})\,, \nonumber \\
a^\dagger_i(\v{k}) &:=& \sqrt{\frac{k}{\hbar\kappa}}\,e^\dagger_i(\v{k}) - \frac{\i}{\sqrt{\hbar\kappa k}}\,K^\dagger_i(\v{k})\,, \nonumber 
\eea
such that
\be
\Big[a_i(\v{k}),a^\dagger_j(\v{k}')\Big] = \delta_{ij}\delta_{\v{k},\v{k}'}\,.
\ee
The one-graviton state with polarization $i$ and momentum $\v{k}$ reads
\be
\Psi_{i,\v{k}}[e_i(\v{k})] = a^\dagger_i(\v{k})\Psi[e_i(\v{k})] 
= 2\sqrt{\frac{k}{\hbar\kappa}}\,e^*_i(\v{k})\,\Psi[e_i(\v{k})] 
\ee
As usual, we can write this also with tensors:
\be
\Psi_{i,\v{k}}[e^{\sss \rm red}_{ab}(\v{k})]
= 2\sqrt{\frac{k}{\hbar\kappa}}\,\epsilon^{ab}_i(\v{k})e^{\sss \rm red*}_{ab}(\v{k})\,\Psi[e^{\sss \rm red}_{ab}(\v{k})] 
\label{gravitonstateFourier}
\ee
The canonical transformation to Ashtekar-Barbero variables adds the phase factor  
\be
\exp\left(\frac{2\i}{\hbar\kappa\beta}\sum_{\v{k}}\,k\,e^*_1(\v{k})e_2(\v{k})\right)\,.
\ee
Next we extend the functional from the reduced to the full configuration space, as we did in sec.\ \ref{From_reduced_to_full_configuration_space}. This gives us
\bea
\Psi_{i,\v{k}}[E^a_l(\v{k})] &=& 2\sqrt{\frac{k}{\hbar\kappa}}\,\epsilon^{la}_i(\v{k})\,e^*_{la}(\v{k})\,\Psi[E^a_j(\v{k})] \nonumber \\
&=& 2\sqrt{\frac{k}{\hbar\kappa}}\,\epsilon^{la}_i(\v{k})\,\frac{1}{\sqrt{V}}\int\d^3 x\;\e^{\i\v{k}\cdot\v{x}}\,e_{la}(\v{x})\,\Psi[E^a_l(\v{k})]\,.
\eea
We replace $e_{la}(\v{x})$ by $\frac{1}{2}\left(\tilde{g}^{la}_\Lambda(\v{x}) - \delta^{la}\right)$ (using again the argument that higher orders in $e$ can be neglected), and arrive at
\be
\Psi_{i,\v{k}}[E^a_l(\v{k})] =
\sqrt{\frac{k}{\hbar\kappa}}\,\epsilon^{la}_i(\v{k})\,\frac{1}{\sqrt{V}}\int\d^3 x\;\e^{\i\v{k}\cdot\v{x}} \left(\tilde{g}^{la}_\Lambda(\v{x}) - \delta^{la}\right)\Psi[E^a_l(\v{k})]\,.
\ee
We bring this into the complexifier form, make the transition to $\H$ and finally apply the gauge projector. The result is the state
\bea
\Psi_{i,\v{k}}
&=& \N\sum_{S\subset\T^*_{\Lambda}} S(0)\, \sqrt{\frac{k}{\hbar\kappa}}\,\epsilon^{la}_i(\v{k})\,\htil^{la*}_S(\v{k}) \nonumber \\
&& \times\,\exp\Bigg[-\frac{1}{4\hbar\kappa}\int\d^3 x\int\d^3 y\;W_\Lambda(\v{x}-\v{y})
\,\htil^{ab}_S(\v{x})\,\htil^{ab}_S(\v{y}) + \mbox{phase term}\Bigg]\; S^*
\eea
in the gauge-invariant Hilbert space $\H_0$. We define an associated wavefunction
\bea
\Psi_{i,\v{k}}(S) &:=&
S(0)\, \sqrt{\frac{k}{\hbar\kappa}}\,\epsilon^{la}_i(\v{k})\,\htil^{la*}_S(\v{k}) \nonumber \\
&& \times\,\exp\Bigg[-\frac{1}{4\hbar\kappa}\int\d^3 x\int\d^3 y\;W_\Lambda(\v{x}-\v{y})
\,\htil^{ab}_S(\v{x})\,\htil^{ab}_S(\v{y}) + \mbox{phase term}\Bigg]\,,
\eea
and write the state as
\be
\Psi_{i,\v{k}} = \N \sum_{S\subset\T^*_{\Lambda}} \Psi_{i,\v{k}}(S) S^*\,.
\ee
In the same way, we construct multiply excited states. Denote the polarizations and momenta of the gravitons by $i_1,\v{k}_1;\ldots;i_N,\v{k}_N$. Then, the $N$-graviton state
in $\H_0$ becomes 
\bea
\Psi_{i_1,\v{k}_1;\ldots;i_N,\v{k}_N}
&=& \N_{i_1,\v{k}_1;\ldots;i_N,\v{k}_N} \sum_{S\subset\T^*_{\Lambda}} \Psi_{i_1,\v{k}_1;\ldots;i_N,\v{k}_N}\!(S) S^*\,,
\eea
where
\bea
\Psi_{i_1,\v{k}_1;\ldots;i_N,\v{k}_N}(S) &:=&
S(0)\,\left(\prod_{n=1}^N \sqrt{\frac{k_n}{\hbar\kappa}}\,\epsilon^{la}_{i_n}(\v{k}_n)\,\htil^{la*}_S(\v{k}_n)\right) \nonumber \\
&& \times\,\exp\Bigg[-\frac{1}{4\hbar\kappa}\int\d^3 x\int\d^3 y\;W_\Lambda(\v{x}-\v{y})
\,\htil^{ab}_S(\v{x})\,\htil^{ab}_S(\v{y}) + \mbox{phase term}\Bigg]\,.
\eea
The normalization factor $\N_{i_1,\v{k}_1;\ldots;i_N,\v{k}_N}$ depends on the excitation number of each mode.

\section{Semiclassical properties of the vacuum state}
\label{Semiclassical_properties_of_the_vacuum_state}

By construction, the form of the vacuum $\Psi_0$ is similar to that of the
Fock space functional we started from. The exponential is still of a Gaussian type, where now the role of the fluctuation variable is played by the spin networks $S$ and their associated (inverse densitized) metric $\gt^{ab}_S$. As before, these fluctuations are non-locally correlated by the kernel $W_\Lambda$.
It is immediate from \eq{spinnetworkcoefficientwithphasefactor} that for most spin networks the coefficient $\Psi_0(S)$ is exponentially damped. An absolute value of the order 0.1 to 1 is only attained for a relatively small class of spin networks: by analogy with quantum mechanics, we say that these spin networks constitute the ``peak region'' of $\Psi_0$. The ``position'' of the peak itself is given by those spin networks for which $\Psi_0(S)$ is exactly 1.

The state inherits its Gaussian property from the semiclassical peakedness of the original Fock state. This suggests that we interpret the peakedness of $\Psi_0$ as the way in which semiclassicality manifests itself on the level of spin networks: i.e.\ we interpret spin networks in the peak region as semiclassical fluctuations, and the spin networks at the peak position as the classical configuration.

From that point of view, it would be interesting to know where exactly the peak is located; that is, for which spin networks $S$ the coefficient $\Psi_0(S)$ reaches its maximum. Below we analyze this question and try to  estimate the peak position: at first for general values of the cutoff length, and then, in section \ref{Limit_lLambda_ll_lp}, for the limit where $l_\Lambda$ is much smaller than the Planck length.

\subsection{Peak position}
\label{Peak}

We have three length scales: the Planck length $l_p$, the length cutoff $l_{\Lambda} = \pi/\Lambda$ corresponding to the momentum cutoff $\Lambda$, and the size of the 3-torus $L = V^{1/3}$. 
The Fourier coefficient of the two-tangent form factor is
\be
F^{ab}_{v,e}(\v{k}) = \frac{1}{\sqrt{V}}\left(\frac{l_p^2\beta}{4}\right)^2\int_0^1\d s\int_0^1\d s'\; \dot{e}^a(s)\dot{e}^b(s')\delta_{\Lambda}(0)\,\e^{-\i\v{k}\cdot\v{x}_v}\,.
\label{Fouriertwotangentformfactor}
\ee
The factor $\delta_\Lambda(0)$ gives 
\bea
\delta_\Lambda(0) &=& \frac{1}{V}\sum_{k\le\Lambda}1 
\approx \frac{1}{V}\frac{1}{(2\pi/L)^3}\int\limits_{k\le\Lambda}\d^3 k \nonumber \\
&=& \frac{1}{(2\pi)^3}\,\frac{4}{3}\pi\Lambda^3 = \frac{\pi}{6}\frac{1}{l_\Lambda^3}
\eea
In section \ref{From_momentum_cutoff_to_triangulation}, we assumed that the edges of the triangulation are straight and that their lengths are more or less equal to
$l_{\Lambda}$. Hence we can approximate \eq{Fouriertwotangentformfactor} by
\bea
F^{ab}_{v,e}(\v{k}) &\approx& L^{-3/2}\,\frac{\beta^2}{16}\,l_p^4\:l_\Lambda n^a_e\, l_\Lambda n^b_e\,
\frac{\pi}{6}\,\frac{1}{l_\Lambda^3}\,\e^{-\i\v{k}\cdot\v{x}_v} \nonumber \\
&=& \eta\,n^a_e n^b_e\,\e^{-\i\v{k}\cdot\v{x}_v}
\,.
\eea
Here, $\eta$ stands for
\be
\eta := \frac{\pi}{96}\,\beta^2\,L^{-3}\,l_p^4\,l_\Lambda^{-1}
\ee
and $\v{n}_e$ denotes the normalized direction vector of the edge $e$:
\be
\v{n}_e := \frac{\v{e}}{\sqrt{\v{e}^2}}
\ee
The $j_e$-dependence of the $S(0)$-factor in \eq{spinnetworkcoefficientFourier} is polynomial. When determining the peak of the state, we can neglect it relative to the exponential dependence. Therefore, the peak condition becomes
\bea
&& \sum_{k\le\Lambda}\omega(\v{k})\left|\eta\,\sum_v\sum_{\mbox{$\ss e$ {\tiny of} $\ss v$}}\,c_e\,n^a_e n^b_e\,
\e^{-\i\v{k}\cdot\v{x}_v} - \delta^{ab}\,\delta_{\v{k},0}\right|^2 \nonumber \\
&=& \omega_0 \left(\sum_v\sum_{\mbox{$\ss e$ {\tiny of} $\ss v$}}\,\eta\,c_e\,n^a_e n^b_e - \delta^{ab}\right)^2
+ \sum_{0 < k\le\Lambda}\omega(\v{k})\left|\sum_v\sum_{\mbox{$\ss e$ {\tiny of} $\ss v$}}\,\eta\,c_e\,n^a_e n^b_e\,\e^{-\i\v{k}\cdot\v{x}_v}\right|^2 = \;\;\mbox{minimal}\,.
\label{peakcondition}
\eea
It is convenient to introduce the matrix
\be
M^{ab} := \sum_v\sum_{\mbox{$\ss e$ {\tiny of} $\ss v$}}\,\eta\,c_e\,n^a_e n^b_e\,,
\ee
its traceless part 
\be
T^{ab} := M^{ab} - \frac{1}{3}\,M\,\delta^{ab}\,,
\ee
and the function
\be
f^{ab}(\v{x}) := \sqrt{V}\,\sum_v\sum_{\mbox{$\ss e$ {\tiny of} $\ss v$}}\,\eta\,c_e\,n^a_e n^b_e\,\delta(\v{x}-\v{x}_v)\,.
\label{definitionf}
\ee
With this notation, condition \eq{peakcondition} takes the form
\bea
&& \omega_0\left(M^{ab} - \delta^{ab}\right)^2 + \sum_{0 < k\le\Lambda}\omega(\v{k})\,|f^{ab}(\v{k})|^2 \nonumber \\
&=& 3\omega_0\left(\frac{1}{3}M - 1\right)^2 + \omega_0\,T^{ab}T^{ab}
+ \sum_{0 < k\le\Lambda}\omega(\v{k})\,|f^{ab}(\v{k})|^2 = \;\;\mbox{minimal}\,.
\label{peakconditionMTf}
\eea
If we minimize each term separately, we obtain the three conditions \setlength{\jot}{0.3cm}
\bean
{\rm (a)} \quad && \mbox{$T^{ab} = 0$, i.e.\ $M^{ab}$ is isotropic.} \\
{\rm (b)} \quad && M = \sum_v\sum_{\mbox{$\ss e$ {\tiny of} $\ss v$}}\,\eta\,c_e = 3\,. \\
{\rm (c)} \quad && \mbox{$f^{ab}(\v{k}) = 0$ for $0<k\le\Lambda$.}
\eean 
Condition (a) requires that on average the spin network edges are isotropic; (b) requires a certain mean value for $c_e = j_e (j_e + 1)$, and (c) demands (weighted) homogeneity of spin network edges up to the scale $l_\Lambda$.
\setlength{\jot}{0cm} 

Let us first consider (b): we define the mean value of $c_e$ on $\T^*_{\Lambda}$ by
\be
\cb_\Lambda = \frac{1}{N_e}\sum_e\,c_e\,,
\ee
where $N_e$ is the total number of edges of the dual complex $\T^*_{\Lambda}$. Noting that $N_e$ equals twice the total number $N_v$ of vertices of $\T^*_{\Lambda}$, we can write condition (b) as
\be
\cb_\Lambda = \frac{3}{4\eta N_v}\,.
\ee
When a triangulation consists of regular tetrahedrons of side length $a$, the dual edge length $a_*$ and the tetrahedron volume $V_T$ are given by
\be
a_* = \frac{1}{\sqrt{6}}\,a\,,\quad V_T = \frac{\sqrt{2}}{12}\,a^3 = \sqrt{3}\,a_*^3\,.
\ee
The tetrahedrons of our triangulation are very close to being regular, therefore
\be
N_v \approx \frac{L^3}{\sqrt{3}\,l_\Lambda^3}
\ee
and 
\bea
\cb_\Lambda &=& \frac{3}{4}\frac{96}{\pi}\,\beta^{-2}\,\frac{L^3\,l_\Lambda}{l_p^4}\,\frac{\sqrt{3}\,l^3_\Lambda}{L^3} 
\nonumber \\
% &=& \frac{72\sqrt{3}}{\pi}\,\beta^{-2}\left(\frac{l_\Lambda}{l_p}\right)^4\,.
&\approx& 40\,\beta^{-2}\left(\frac{l_\Lambda}{l_p}\right)^4\,.
\eea
If we define the number $j_\Lambda$ by  
\be
j_\Lambda(j_\Lambda + 1) := 40\,\beta^{-2}\left(\frac{l_\Lambda}{l_p}\right)^4\,,
\label{jLambda}
\ee
condition (b) becomes
\be
\cb_\Lambda = j_\Lambda(j_\Lambda + 1)\,.
\label{conditioncb}
\ee 
For certain values of $l_\Lambda$, the ``spin'' $j_\Lambda$ is a half-integer (see \fig{plot}). In these cases, it is easy to satisfy (a), (b) and (c): take the entire dual complex $\T^*_\Lambda$ as the spin network graph and label all edges with the spin $j_\Lambda$. Clearly, the mean value of $c_e$ is $j_\Lambda(j_\Lambda + 1)$, so (b) is fulfilled. Since we have chosen a very homogenuous and isotropic triangulation, the uniform spin distribution over all edges of $\T^*_\Lambda$ automatically implies isotropy (a) and homogeneity (c).
\psfrag{x}{\hspace{-0.5cm}\parbox[t]{1cm}{\hspace{0cm} \\ $\frac{l_\Lambda}{\sqrt{\beta}\,l_p}$}}
\psfrag{j}{\hspace{-0.3cm}\parbox[c]{1cm}{$j_{\sss \Lambda}$ \\ \hspace{0cm}}}
\pic{plot}{Dependence of $j_\Lambda$ on cutoff length $l_\Lambda$.}{6cm}{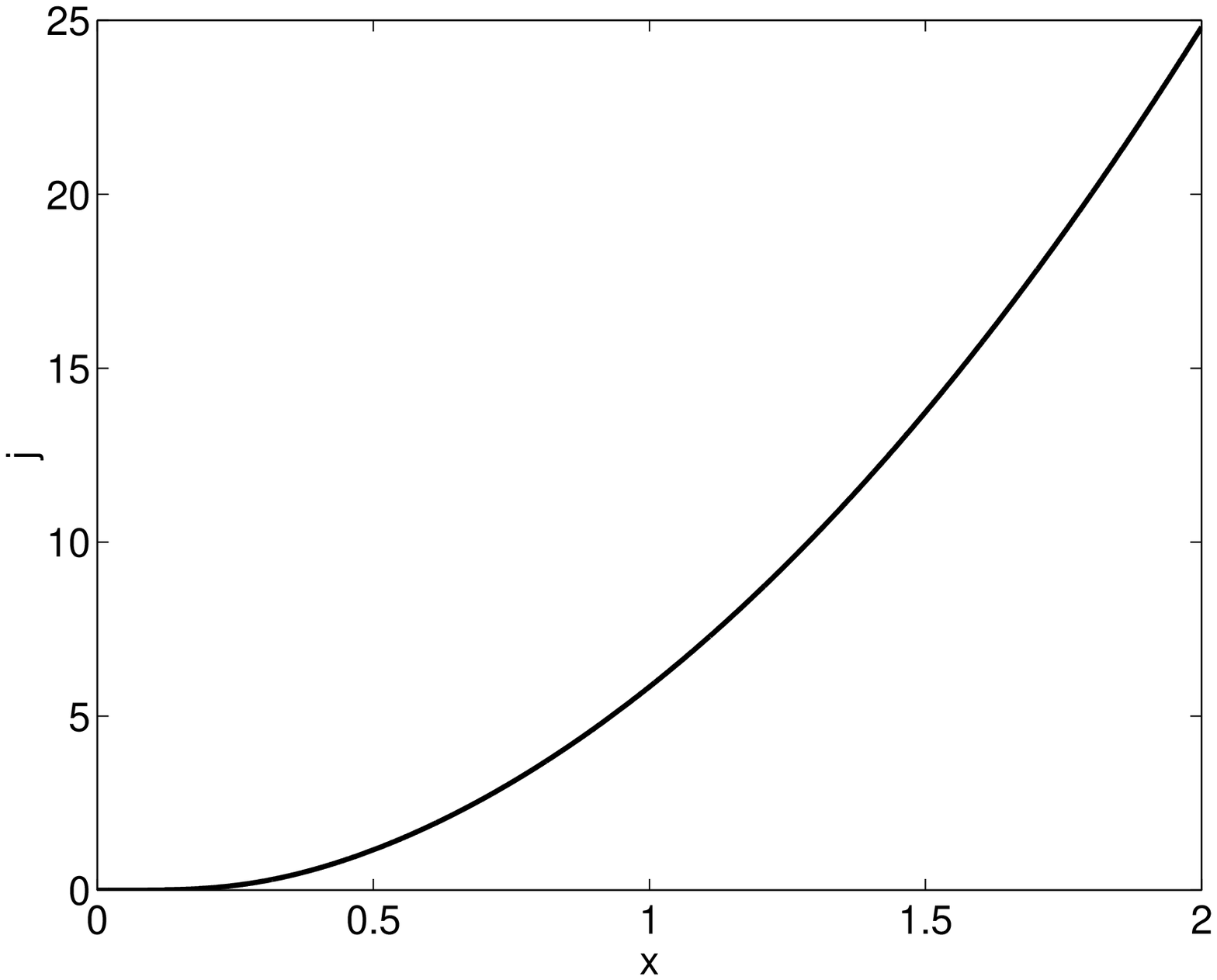}

Thus, for values of $l_\Lambda$ which yield a half-integer $j_\Lambda$, the peak of the state \eq{spinnetworkcoefficientwithphasefactor} lies at spin networks which have the entire dual complex as graph and all spin labels equal to $j_\Lambda$. This peak exhibits a degeneracy due to the remaining freedom in choosing intertwiners at vertices.  We see from \eq{jLambda} that the required spin depends strongly on the length cutoff $l_\Lambda$: 
beyond $l_\Lambda > 0.7\sqrt{\beta}\,l_p$ it grows quadratically in $l_\Lambda$. In the range $0.4\sqrt{\beta}\,l_p < l_\Lambda < 0.7\sqrt{\beta}\,l_p$, it takes values between 1/2 and 10.

What are the peaks of the state when $j_\Lambda$ lies between two half-integers $j_1 < j_2$? In that case, we have two opposing tendencies: condition (b) requires a mean value $\cb_\Lambda = j_\Lambda(j_\Lambda + 1)$ between $j_1(j_1+1)$ and $j_2(j_2+1)$, and therefore at least an alternation of labels between $j_1$ and $j_2$. This is in conflict with condition (c), which demands spatial homogeneity on all scales larger or equal to $l_\Lambda$. That is, we cannot minimize all terms in \eq{peakcondition} simultaneously.

There are essentially two possibilities how the peak state could behave: one is that it becomes some blend-over of the uniform spin distributions for $j_\Lambda = j_1$ and $j_\Lambda = j_2$. The other possibility: the configuration keeps the uniform spin $j_1$ while $j_\Lambda$ moves away from $j_1$ towards $j_2$, and at some point it makes an abrupt transition to the uniform $j_2$-configuration.

\subsection{Limit $l_\Lambda \ll l_p$}
\label{Limit_lLambda_ll_lp}

Instead of discussing this in detail for general $l_\Lambda$, we will concentrate on the most interesting case, namely, when $l_\Lambda$ is much smaller than the Planck length. Then, $j_\Lambda$ lies between 0 and 1/2. To determine the peak states, we introduce three classes of spin networks that we call $S_b$, $S_{b\!\!\!\slash}$ and $S_{ab\v{c}}$. 

$S_b$ denotes the class of all spin networks on $\T^*_\Lambda$ that meet condition (b), whereas
$S_{b\!\!\!\slash}$ stands for those which violate (b). $S_{ab\v{c}}$ consists of the spin networks which satisfy (a) and (b), and minimize the violation of condition (c), i.e.\ they minimize the third term in \eq{peakconditionMTf}, while the first two terms are zero. 

We will first determine the properties of spin networks in $S_{ab\v{c}}$, and then argue that this class of states is the minimizing solution for condition \eq{peakconditionMTf}. 

We begin by analyzing condition (b).
Suppose that for some $\Lambda'<\Lambda$, there is a triangulation of the type $\T_{\Lambda'}$
whose dual complex $\T^*_{\Lambda'}$ is coarser than $\T^*_\Lambda$ (for the definitons, see p.\ \pageref{definitionofdualcomplex}). Consider the restricted mean value 
\be
\cb_{\Lambda'} := \frac{1}{N_{e'}}\sum_{e'}\,c_e\,,
\ee
where the sum runs only over edges of the coarser complex $\T^*_{\Lambda'}$, and $N_{e'}$ is the total number of its edges. By repeating the steps that led up to \eq{conditioncb}, we find that for spin networks on $\T^*_{\Lambda'}$, condition (b) is equivalent to
\be
\cb_{\Lambda'} = j_{\Lambda'}(j_{\Lambda'} + 1)\,,
\label{generalizedconditioncb}
\ee
This equation generalizes equation \eq{conditioncb} in the sense that it expresses (b) as a constraint on various scales of $\T^*_\Lambda$. In particular, for uniform spin distributions, it gives a relation between spin and scale of the distribution. 

Using this reformulation of condition (b), we immediately find a large class of spin networks that solve it:
according to \eq{generalizedconditioncb}, a spin network meets condition (b), if it has a dual
complex $\T^*_{\Lambda'}\subset\T^*_\Lambda$ as its graph and a uniform spin labelling with $j_{\Lambda'}$. 
For given $j_{\Lambda'}$, we call the set of such spin networks $S_{j_{\Lambda'}}$.
The admissible scales $\Lambda'$ are those where $j_{\Lambda'}$ takes a half-integer value.
For example, $j_{\Lambda'} = 1/2$ corresponds to the scale $0.4\sqrt{\beta}\,l_p$.
By definition, a complex $\T^*_{\Lambda'}$ is nearly isotropic, so spin networks in $S_{j_{\Lambda'}}$ meet also condition (a).

The third condition on $S_{ab\v{c}}$ states that the spin networks should minimize the term
\be
\I = \sum_{0 < k\le\Lambda}\omega(\v{k})\,|f(\v{k})|^2\,,
\label{thirdterm}
\ee
which measures inhomogeneity. A state in the class $S_{j_{\Lambda'}}$ breaks homogeneity (condition (c)) at the scale $\Lambda'$. Using \eq{definitionf} and \eq{jLambda}, we see that it yields 
\bea
\I &=& \sum_{0 < k\le\Lambda}\omega(\v{k})\left|\sum_v\sum_{\mbox{$\ss e$ {\tiny of} $\ss v$}}\,\eta\,c_e\,n^a_e n^b_e\,\e^{-\i\v{k}\cdot\v{x}_v}\right|^2 \sim 
\Lambda'\,\Big[\eta\,j_{\Lambda'}(j_{\Lambda'} + 1)\,N_{v'}\Big]^2  \nonumber \\
&\sim & 
\Lambda'\,\left[\eta\,j_{\Lambda'}(j_{\Lambda'} + 1)\left(\frac{l_\Lambda}{l_{\Lambda'}}\right)^3\!\! N_v\right]^2  
\nonumber \\
&\sim & 
\Lambda'\,\left[j^{-1}_\Lambda(j_\Lambda + 1)^{-1}\,j_{\Lambda'}(j_{\Lambda'} + 1)\left(\frac{l_\Lambda}{l_{\Lambda'}}\right)^3\right]^2  \nonumber \\
&\sim & 
\Lambda'\,\left[\left(\frac{l_\Lambda}{l_p}\right)^{-4}\,\left(\frac{l_{\Lambda'}}{l_p}\right)^4\,
\left(\frac{l_\Lambda}{l_{\Lambda'}}\right)^3\right]^2  \nonumber \\
&=& \Lambda'\left(\frac{l_{\Lambda'}}{l_\Lambda}\right)^2 \sim \frac{\Lambda^2}{\Lambda'}\,.
\label{HforSjLambda}
\eea
In going from the third to the fourth line, we used that $\eta N_v \sim j^{-1}_\Lambda(j_\Lambda + 1)^{-1}$.
Loosely speaking, high momentum wins over small momentum scale, and of all the sets $S_{j_{\Lambda'}}$ it is the one with the highest momentum scale $\Lambda'$ that minimizes \eq{thirdterm}. Clearly, that set is $S_{\sss 1/2}$. 

Given the domination of low spin in \eq{HforSjLambda}, it appears unlikely that configurations with mixed spin labels lead to the same or a lower value. If that expectation is correct, $S_{\sss 1/2}$ does not only minimize the sequence $S_{\sss 1/2}, S_{\sss 1}, S_{\sss 3/2}, \ldots$, but also all other states that satisfy (a) and (b). That is, $S_{\sss 1/2}$ equals $S_{ab\v{c}}$. (When saying so, we are slightly imprecise: 
the minimizing conditions do not require that the graph takes {\it exactly} the form of a dual complex of a triangulation. The set $S_{ab\v{c}}$ can also contain spin networks with other graphs $\G_{\Lambda'}$ as long as they are isotropic, homogeneous and have a length scale corresponding to $j_{\Lambda'} = 1/2$.
When speaking of $S_{\sss 1/2}$ in the following, we mean to include these additional spin networks.)

Let us now consider the relation of $S_{ab\v{c}}$ to the entirety of configurations on $\T^*_\Lambda$: 
we distinguish between configurations that satisfy (b) (the class $S_b$) and those which do not (the class $S_{b\!\!\!\slash}$). By construction, $S_{ab\v{c}}$ is the subset of $S_b$ that comes closest to the minimum.
Regarding $S_{b\!\!\!\slash}$, there is, in principle, the possibility that the violation of (b) is compensated by greater homogeneity of the spin distribution. To check this, we compare expression \eq{peakconditionMTf} for 
$S_{ab\v{c}}$ and $S_{b\!\!\!\slash}$: for $S_{ab\v{c}} = S_{\sss 1/2}$ we just get the inhomogeneity, i.e.\
\be
\I \sim \Lambda_p\left(\frac{l_p}{l_\Lambda}\right)^2 \sim \frac{l_p}{l^2_\Lambda}\,.
\label{inhomogeneityShalf}
\ee
On the side of $S_{b\!\!\!\slash}$, we obtain
\bea
\lefteqn{3\omega_0\left(\frac{1}{3}M - 1\right)^2 + \omega_0\,T^{ab}T^{ab}
+ \sum_{0 < k\le\Lambda}\omega(\v{k})\,|f(\v{k})|^2} \nonumber \\
&\ge & 3\omega_0\left(\frac{1}{3}\sum_v\sum_{\mbox{$\ss e$ {\tiny of} $\ss v$}}\,\eta\,c_e - 1\right)^2 \nonumber \\
&\sim & \omega_0\left(\frac{4}{3}\,\eta\,N_v\,\cb_\Lambda - 1\right)^2 \nonumber \\
&\sim & \omega_0\,\frac{1}{j^2_\Lambda(j_\Lambda + 1)^2}\left[\cb_\Lambda - j_\Lambda(j_\Lambda + 1)\right]^2 
\label{fixedrelativedifference} \\
&\sim & \omega_0\left(\frac{l_p}{l_\Lambda}\right)^8\left[\cb_\Lambda - j_\Lambda(j_\Lambda + 1)\right]^2\,.
\label{fixeddifference}
\eea
If we think about states in $S_{b\!\!\!\slash}$ whose mean value $\cb_\Lambda$ has a finite difference to $j_\Lambda(j_\Lambda + 1)$, the term \eq{fixeddifference} blows up much faster than 
\eq{inhomogeneityShalf} as $l_\Lambda\to 0$, so the $S_{\sss 1/2}$ states are preferred. 
Alternatively, we could consider states $S^r_{b\!\!\!\slash}$ that have a fixed {\it relative} difference between $\cb_\Lambda$ and $j_\Lambda(j_\Lambda + 1)$, so that \eq{fixedrelativedifference} remains constant. In that case, the inhomogeneity becomes the dominating criterion. Since $j_\Lambda(j_\Lambda + 1)$ gets close to zero, $\cb_\Lambda$ does so too, and the states $S^r_{b\!\!\!\slash}$ become similar to $S_b$ states. In fact, there is no obvious reason why such states should have advantage over the $S_j$-states. Hence we expect again that the $S_{\sss 1/2}$ class provides the lowest value.

This would mean that for $l_\Lambda \ll l_p$, the peak of the state consists of spin networks with the following properties:  they are uniformly labelled by 1/2, their graphs lie on $\T^*_\Lambda$, are homogeneous and isotropic, and the effective length scale of the graphs is close to $\sqrt{\beta}\,l_p$, regardless of how fine $\T^*_\Lambda$ is.

\section{Summary and discussion}
\label{Summary_and_discussion}

\subsection{Summary of results}

The basic idea of our approach is to start from the free Fock vacuum of linearized gravity and construct from it a state $\Psi_0$ that could play the role of the ``free'' vacuum in loop quantum gravity. In making the transition from Fock to loop state, we have to take various choices that relate the field variables of the former to the polymer-like degrees of freedom of the latter. We have done so with the aim of making the state reasonably simple while preserving, as far as possible, the physical properties of the original state.

Let us repeat the logic of our construction: 
\begin{enumerate}
\item
We linearize extended ADM gravity around a flat background on $T^3\times \bR$.
\item We apply a reduced phase space quantization and specify the free vacuum.
\item We perform the canonical transformation to linearized Ashtekar-Barbero variables and implement it as a unitary transformation in the quantum theory.
\item We regularize the state by a momentum cutoff $\Lambda$.
\item The transition to the LQG Hilbert space is achieved in five steps:
\begin{enumerate}
\item We extend the state from a functional of reduced to a functional of full triads.
\item We replace the fluctuation $e^a_i = E^a_i - \delta^a_i$ by 
$$\frac{1}{2}\left(\tilde{g}^{ab}(\v{x}) - \delta^{ab}\right) \equiv \frac{1}{2}\left(E^a_i(\v{x})E^b_i(\v{x}) - \delta^{ab}\right)\,,$$
the fluctuation in the densitized inverse metric associated to $E^a_i$.
\item We bring the state into the complexifier form.
\item We replace the momentum cutoff $\Lambda$ by a regular triangulation $\T_\Lambda$ of length scale $\pi/\Lambda$, and thus obtain a functional of generalized connections.
\item Gauge projection yields a state in the gauge-invariant Hilbert space $\H_0$.
\end{enumerate}
\end{enumerate}
Ideally, this procedure should be completed by an averaging over the 3d-diff group, so that one receives a state in $\H_{\rm diff}$. We have not carried out this step.

The state we get is a superposition 
\be
\Psi_0 = \N \sum_{S\subset\T^*_{\Lambda}}\Psi_0(S)S^*\,.
\label{superpositionofspinnetworks}
\ee
The sum ranges over all spin networks $S$ whose graph lies on the dual complex $\T^*_{\Lambda}$ of the triangulation. The coefficients $\Psi_0(S)$ are given by
\be
\Psi_0(S) = S(0)
\exp\Bigg[-\frac{1}{4\hbar\kappa}\int\d^3 x\int\d^3 y\;W_\Lambda(\v{x}-\v{y})
\,\htil^{ab}_S(\v{x})\,\htil^{ab}_S(\v{y}) + \mbox{phase term}\Bigg]\,.
\label{spinnetworkcoefficientagain}
\ee
To each spin network $S$ we associate an inverse metric, and $\htil^{ab}_S$ stands for the difference between $\gt^{ab}_S$ and the flat background metric:
\be
\tilde{h}^{ab}_S = \gt^{ab}_S - \delta^{ab}\,.
\ee
$\gt^{ab}_S$ is distributional and has only support on vertices of the spin network graph:
\be
\gt^{ab}_S(\v{x}) = \sum_v\,\gt^{ab}_S(v)\,\delta(\v{x}-\v{x}_v)
\ee
and
\be
\gt^{ab}_S(v) = \frac{\beta^2}{16}\,l_p^4\delta_\Lambda(0)\;\sum_{\mbox{$\ss e$ {\tiny of} $\ss v$}}j_e(j_e+1)\int_0^1\d s\int_0^1\d s'\;\;\dot{e}^a(s)\dot{e}^b(s')\,.
\label{discretemetricassociatedtoS}
\ee
The functions $W_\Lambda(\v{x}-\v{y})$ and $\delta_\Lambda(\v{x})$ are regularized forms of the kernel $1/|\v{x}-\v{y}|^4$ and the delta distribution respectively.
On a hypercubic lattice, the phase term takes the form
\be
\frac{\i}{4\hbar\kappa\beta}\sum_v 
\,\htil^{ab}_S(v)
\,\varepsilon^{acd}\,\nabla^c
\htil^{ab}_S(v)
\ee 
where $\nabla^c$ denotes the lattice derivative. Here, we are working with triangulations, so this formula has to be adapted to the geometry of the triangular lattice. In complete analogy, we have also constructed $N$-graviton states.

In sec.\ \ref{Semiclassical_properties_of_the_vacuum_state}, we investigated the $S$-dependence of the coefficient $\Psi_0(S)$: it is of Gaussian type and exponentially damped for most spin networks. 
At the peak of the Gaussian, the labelling is determined by a characteristic ``spin'' $j_\Lambda$ which depends strongly on the cutoff scale (see \fig{plot}):
\be
j_\Lambda(j_\Lambda + 1) = 40\,\beta^{-2}\left(\frac{l_\Lambda}{l_p}\right)^4\,,
\label{jLambdaagain}
\ee
We have shown that for cutoff lengths $l_\Lambda$ where $j_\Lambda$ is a half-integer, the peak spin networks have the entire dual complex $\T^*_\Lambda$ as their graph and the edges are uniformly labelled by $j_\Lambda$. For values of $l_\Lambda$ where $j_\Lambda$ lies between two nonzero half-integers $j_1 < j_2$, the graph is again the entire complex and we expect that the spin labelling is either uniformly $j_1$,  $j_2$, or a mixture of both. When $l_\Lambda$ is much smaller than the Planck length, 
our analysis supports the idea that the peak spin networks become independent of the cutoff, have spin label 1/2 and homogenous isotropic graphs at a length scale close to $\sqrt{\beta}\,l_p$.

\subsection{Relation to other approaches}

Let us point out some similarities and differences to what has already appeared in the literature:

The idea of using linearized states is not new and has been investigated before by Ashtekar, Rovelli \& Smolin \cite{ARSgravitonsloops} and Varadarajan \cite{V}. In both cases, however, the Fock states were transferred to a loop representation of {\it linearized} gravity, and not to the full representation, as we do here.
Our state is also related to the class of complexifier coherent states which are used by Thiemann and collaborators \cite{TGCS1}-\cite{SaT2}: what distinguishes our proposal is the particular choice of ``complexifier'' and the fact that we leave the background field outside the delta functional --- a circumstance which facilitates the investigation of peak properties considerably. One could say that we keep the background in the ``complexifier'', though in the strict sense of the word this object is no complexifier anymore. Thiemann also introduces cutoff states that have support on a fixed graph, but it is not clear which graph should be chosen or how such a choice should be justified. In our case, the cutoff graph is fixed on physical grounds, namely, by the requirement that it emulates the momentum cutoff of the original Fock state.

Ashtekar and Lewandowski \cite{AL} have proposed a coherent state that is similar to ours in that it is based on $J$-insertions at vertices (see eqn.\ \eq{smearedmetricoperatoronStwithcrossterms}). The way in which the $J$-operators (and form factors) are contracted is different, however. Moreover, their convolution kernel is $1/k$ (up to regulators), while ours corresponds to $k$. This difference can be explained as follows: Ashtekar and Lewandowski start out from $U(1)$ Maxwell theory, where the Hamiltonian is $H \sim E^2 + k^2 B^2$, and then generalize from $U(1)$ to $SU(2)$. In the $E$-representation, the vacuum wavefunctional contains the kernel $1/k$ and that is also the kernel which appears in the complexifier. Our state, on the other hand, comes from the linearization of gravity where the linearized Hamiltonian reads $H \sim K^2 + k^2 e^2$. Thus, the vacuum functional in $e$ has $k$ as its kernel and inherits it to the loop state. 

Our results on the peak of the state clearly show a connection to the early weave approach \cite{ARSweaves}-\cite{Iw}, where a smeared metric operator was used to define eigenstates of the 3-metric. The construction of $\Psi_0$ involves a smeared inverse metric operator
\be
\hat{\tilde{g}}^{ab}_{\Lambda} \equiv 
\int\d^3 x'\;\hE^a_i(\v{x})\delta_{\Lambda}(\v{x}-\v{x}')\hE^b_i(\v{x}')\,,
\label{smearedmetricoperatoragain}
\ee
and one of the conditions for maximizing $\Psi_0(S)$ consists in the requirement that $S$ is an  eigenstate of \eq{smearedmetricoperatoragain} with eigenvalue equal to $\delta^{ab}$. Thus, the peak spin networks of $\Psi_0$ can be considered as a certain form of weave. For $l_\Lambda > \sqrt{\beta}\,l_p$, these weave states have the entire dual complex as graph, and the cutoff-dependence of the spin labelling follows the same logic as in the weave approach: for a large cutoff length $l_\Lambda$, the edges of the dual complex are dispersed far apart and they have to be labelled with large spins to generate a field strength equivalent to the background. For smaller $l_\Lambda$, the available edges are more densely distributed and a smaller mean spin suffices to attain the same field strength. It is somewhere near $l_\Lambda = \sqrt{\beta}\,l_p$ where the mean spin reaches values of the order 1. 

If $l_\Lambda$ is considerably smaller than the Planck length, one would expect that the average $j_e$ has to be much smaller than the minimum nonzero spin 1/2, so that only now and then an edge carries a spin 1/2, while most other edges are labelled trivially. In other words, one could choose the triangulation increasingly finer than the Planck length, but the spin networks would stay apart as if they were trying to keep a triangulation fineness at a larger length scale. This is, in fact, what our analysis indicates: namely, that for $l_\Lambda \ll l_p$, the peak spin networks are in the fundamental representation and have isotropic homogenuous graphs whose edges maintain a distance scale close to the Planck scale. This stands in correspondence to the key result on weaves \cite{ARSweaves}, which states that the approximation of 3-metrics cannot be improved by making the lattice finer than $l_p$. Let us mention that in \cite{CRe}, Corichi and Reyes have also constructed a state that is peaked around weave-like spin networks in the fundamental representation.

\subsection{Continuum limit of free vacuum?}
\label{Continuum_limit_of_free_vacuum}

The cutoff-independence of the peak suggests that the entire state
could become independent of $l_\Lambda$ when $l_\Lambda$ goes to zero. Note that the limit should be taken as
\be
\lim\limits_{\Lambda\to\infty} \N_\Lambda \sum_{S\subset\T^*_{\Lambda}}\Psi_0(S)S^* 
\ee
where $\N_\Lambda$ normalizes the state for each value $l_\Lambda$ in the sequence. Since $\gt^{ab}_S$ diverges for $l_\Lambda\to 0$, any deviation from the peak will be infinitely suppressed: the limiting state loses its spreading and becomes effectively a weave state. 

The reason for this behaviour lies in the fact the we have not replaced the momentum cutoff completely when going from the Fock to the loop state. Instead we ended up with a mixed scheme where part of the regularization is provided by the cutoff graph $\T^*_{\Lambda}$ and part of it by the momentum cutoff in the kernel $W_\Lambda$ and the factor $\delta_\Lambda(0)$ in \eq{discretemetricassociatedtoS}. Such a mixed regularization may be ok for $l_\Lambda > l_p$, but it becomes physically questionable when the regulator is removed:
then, the triangulation part leads to an effective cutoff at the Planck scale (``space is discrete''), while the momentum part lets $\delta_\Lambda(0)$ diverge (``space is continuous'').

To arrive at a more meaningful continuum limit, it would be desirable to express the regularization entirely in terms of the triangulation. Carlo Rovelli has suggested that one might achieve this by starting from a modification of the Fock state \eq{complexifierform}: observe that the kernel $W_\Lambda$ and the delta function $\delta_\Lambda$ require the background triad $E_{cl}$ as an input, so we write them as $W_{\Lambda,E_{cl}}$ and $\delta_{\Lambda,E_{cl}}$. In the state functional, we replace 
these quantities by $W_{\Lambda,E}$ and $\delta_{\Lambda,E}$, since it does not change the part of the exponential which is quadratic in the fluctuation $E-E_{cl}$. 
The kernel $W_{\Lambda,E}$ is the same as $\sqrt{-\Delta_{\Lambda,E}}$ where $\Delta_{\Lambda,E}$ denotes the $\Lambda$-regularized Laplace operator associated to the triad $E$. 

The next step is the transition to the LQG state, that is, one has to decide on how
to implement the propagator $W_{\Lambda,E} = \sqrt{-\Delta_{\Lambda,E}}$ in the LQG
complexifier. The first thing to note is that when making this transition,
the $\Lambda$ in $\Delta_{\Lambda,E}$ can be dropped because the triangulation
takes care of the momentum regularization. Therefore, what remains to be done
is to find a loop quantization $\Delta_{\hE}$ of $\Delta_E$ on this
triangulation. If things are simple, it will act diagonally on spin networks
and contributes a factor $W_{e,e'}(S)$ for every pair $e, e'$ of edges of
 the spin network $S$ on which the complexifier acts. In a similar way, one might 
 be able to define a loop analogue of $\delta_\Lambda(0)$.

The conjecture is that by this procedure one arrives at a ``free'' vacum that
becomes independent of the cutoff when $l_\Lambda$ goes to zero and {\it keeps} the form
that it had for $l_\Lambda \sim l_p$.

\subsection{Free theory, perturbation theory and renormalization?}
\label{freetheoryperturbationtheoryandrenormalization}

We finally come to the big question that motivates this work: can one obtain a semiclassical perturbation series for loop quantum gravity, and if yes, how?

To start with, let us recall how perturbation theory works in ordinary QFT, or even simpler, in quantum mechanics. Consider the anharmonic oscillator with the Hamiltonian
\be
H = \frac{1}{2}P^2 + \frac{1}{2}\omega^2 (Q-Q_0)^2 + \lambda (Q-Q_0)^4\,.
\ee
Assume that we want to do perturbation theory around the static solution $Q(t) = Q_0$, $P(t) = 0$. For that purpose, we introduce the relative variables
\be
q := Q - Q_0\,,\quad p := P\,,
\ee
and ``expand'' the Hamiltonian in $q$ and $p$:
\be
H = \frac{1}{2}p^2 + \frac{1}{2}\omega^2 q^2 + \lambda q^4\,.
\ee
Perturbation theory rests on the idea that we consider only processes that involve a certain subset of states; namely those whose wavefunction is peaked near the classical position and momentum, and strongly damped farther away from it. If the width of these states is sufficiently small, one may ignore higher order terms to first approximation, and use instead the free Hamiltonian
\be
H_0 = \frac{1}{2}p^2 + \frac{1}{2}\omega^2 q^2\,.
\ee
The associated free vacuum is
\be
\psi_0(q) = \N\exp\left(-\frac{\omega}{2\hbar} q^2\right)\,.
\ee
For this approximation to be consistent, the width of $\Psi_0$ should be small enough that the neglection of higher orders is admissible. We check this by applying the full potential $\hat{V}$ to $\psi_0$:
\be
\hat{V}\psi_0(q) = \N\left(\frac{1}{2}\omega^2 q^2 + \lambda q^4\right)\exp\left(-\frac{\omega}{2\hbar} q^2\right)\,.
\label{consistencycheck}
\ee
The width of $\psi_0$ is $\sqrt{\hbar/\omega}$, so the consistency condition becomes
\be
\frac{1}{2}\omega^2\frac{\hbar}{\omega} \gg \lambda\left(\frac{\hbar}{\omega}\right)^2\,,
\ee
or
\be
\lambda \ll \frac{\omega^3}{2\hbar}\,.
\ee
If it is satisfied, there is hope that a perturbative treatment leads to meaningful results. Otherwise we are in the non-perturbative regime.

Let us transfer this logic to the linearization of ADM gravity on $\T^3\times\bR$. When we keep all orders in the reduced fluctuation variables, the Hamiltonian reads
\be
H = \frac{1}{\kappa}\int\d^3 x\left[\left(K^l_aK^j_b - K^j_aK^l_b\right)\left(\delta^a_j + e^a_j\right)\left(\delta^b_l + e^b_l\right) - \left|\det\left(\delta^a_j + e^a_j\right)\right|\,R\left(\delta^a_j + e^a_j\right)\right]\,. 
\label{Hamiltonianallorders}
\ee
Suppose we quantize this system as we did in the free case, i.e.\ by a standard canonical quantization and using a momentum cutoff $\Lambda$ as a regulator. Then, a calculation analogous to \eq{consistencycheck} yields the expansion parameter
\be
\alpha := \frac{\sqrt{\hbar\kappa_{\sss \Lambda}}\,L^{3/2}}{l^{5/2}_\Lambda}\,.
\label{expansionparameter}
\ee
Here, we have taken $\kappa_{\sss \Lambda}$ to be the gravitational coupling at the cutoff scale. $L$ is the size of the torus and the time scale of the process we consider. The simplest way to relate $\kappa_{\sss \Lambda}$ to the classical value is dimensional scaling, i.e.\
\be 
\kappa_{\sss \Lambda} = \frac{l_\Lambda^2}{L^2}\,\kappa\,.
\ee
In that case, the parameter \eq{expansionparameter} becomes
\be
\alpha = \frac{l_p\,L^{1/2}}{l_\Lambda^{3/2}}\,.
\label{secondexpansionparameter}
\ee 
This estimate indicates that for $L$ of the order $1{\rm cm}$, perturbation theory requires a cutoff length larger than $10^{-22}{\rm cm}$. For $l_\Lambda = l_p$, the theory would be highly non-perturbative.

Of course, what we actually want to analyze is loop quantum gravity, and not a conventional quantization of gravity with an arbitrary cutoff. With that aim in mind, we transferred the free vacuum of \eq{Hamiltonianallorders} to the LQG Hilbert space. The semiclassical spreading around $e = 0$ turned into a Gaussian spreading around weave-like spin networks. The more the spin labelling of a spin network differs from that of the weave, the more it is damped in the sum \eq{superpositionofspinnetworks}. The quantity that measures the deviation from the peak is 
\be
\tilde{h}^{ab}_S = \gt^{ab}_S - \delta^{ab}\,,
\ee
and stands in obvious correspondence to the fluctuation $e$ of the traditional approach.
Therefore, it appears natural to ask the following questions: could this correspondence give us a hint on how semiclassicality manifests itself in full loop quantum gravity? Could we take the size of $\tilde{h}^{ab}_S$, or a similar quantity, as the loop analogue of the fluctuation, and use it as a measure for telling whether a spin network is a ``small'' or a ``large'' fluctuation w.r.t.\ a semiclassical state? Suppose that we let the full Hamiltonian constraint $\hC$ act on a state like $\Psi_0$ that is peaked around weaves. Would there be parts of $\hC$ that are more relevant than others, in the same way that $\omega^2 q^2/2$ is more relevant than $\lambda q^4$ when acting with the potential on $\exp(-\omega q^2 / 2\hbar)$?
In other words, can we write $\hC$ as a sum
\be
\hC = \hC_0 + \hC_1 
\ee
such that to first approximation $\hC_1 \Psi_0$ can be neglected relative to $\hC_0 \Psi_0$?
To answer that question, one would have to analyze if the quantity
\be
\Psi_0(S)\,\hC S 
\ee
can be approximated for spin networks close to the peak weaves. I.e.\ for those spin networks whose spin assignments differ from the weaves on only few edges and only by little spin. 

Thus, we are led to an ansatz that appears very similar to the one used by Smolin when he analyzes string perturbations of causal spin networks \cite{Sstring}: in that case, the ``differing'' edges are chosen to form loops and are specified on an entire history of causal spin networks. The sequence of loop-like deviations is viewed as a string worldsheet, and the resulting variation in the amplitude is to some extent evaluated.

The inherent cutoff-property of LQG suggests that loop quantization is related to conventional quantization schemes where a cutoff $l_\Lambda = l_p$ is put in by hand. Similarly as there, the expansion parameter could be large and perturbation theory impossible. In that case, a coarse-graining procedure \cite{Ma,O} may be necessary to compute an effective Hamiltonian constraint for a lower scale $l_\Lambda$, where the expansion parameter is small. Our state $\Psi_0$ might be useful in such attempts because it is an element in the full Hilbert space and represents at the same time a semiclassical state at the cutoff scale $l_\Lambda$. One could try to extract an effective matrix element 
\be
\b \Psi_0 | P_\Lambda | \Psi_0 \k\quad\mbox{from}\quad\b \Psi_0 | P | \Psi_0 \k\,, 
\ee
where $P$ is the ``bare'' projector onto the physical Hilbert space, and $P_\Lambda$ is based on a renormalized Hamiltonian constraint at the cutoff scale $l_\Lambda < l_p$.

Admittedly, the considerations of this last section are vague and need to be concretized in many ways. We hope to do so in future work.

\subsection*{Acknowledgements}
I thank Carlo Rovelli, Thomas Thiemann, Laurent Freidel and Abhay Ashtekar for helpful discussions and suggestions. 
I also thank the members of the CPT, Marseille for their hospitality. 
This work was supported by the Daimler-Benz foundation and DAAD.


\begin{thebibliography}{99}

%%%%%%%%%%%%%%% introductions %%%%%%%%%%%%%%%
\bibitem{T}
T.\ Thiemann, {\it Introduction to Modern Canonical Quantum General
Relativity}, gr-qc/0110034.
\bibitem{R}
C.\ Rovelli, ``Quantum Gravity'' (Cambridge University Press, Cambridge 2004), for a draft see
http://www.cpt.univ-mrs.fr/\symbol{126}rovelli/.
\bibitem{B}
J.C.\ Baez, {\it An introduction to spin foam models of quantum gravity and BF theory}, Lect.\ Notes Phys. {\bf 543}, 25 (2000), gr-qc/9905087.
\bibitem{P}
 A.\ Perez, {\it Spin foam models for quantum gravity}, Class.\ Quant.\ Grav.\ {\bf 20}, R43 (2003), gr-qc/0303026.

%%%%%%%%%%%%%%% Kodama state %%%%%%%%%%%%%%%
\bibitem{S} 
L.\ Smolin, {\it Quantum gravity with a positive cosmological constant}, hep-th/0209079.
\bibitem{FS}
L.\ Freidel, L.\ Smolin, {\it The linearization of the Kodama state}, hep-th/0310224.
\bibitem{St} 
A.\ Starodubtsev, {\it Topological excitations around the vacuum of quantum gravity.\ 1.\ The symmetries of the vacuum}, hep-th/0306135.
\bibitem{SSo}
L.\ Smolin, C.\ Soo, {\it The Chern-Simons invariant as the natural time variable for classical and quantum cosmology}, Nucl.\ Phys.\ B {\bf 449}, 289 (1995), gr-qc/9405015.
\bibitem{So}
C.\ Soo, {\it Wave function of the universe and Cherns-Simons perturbation theory}, Class.\ Quant.\ Grav.\ {\bf 19}, 1051-1064 (2002), gr-qc/0109046. 
\bibitem{SoCh}
C.\ Soo, L.N.\ Chang, {\it Superspace dynamics and perturbations around 'emptiness'}, Int.\ J.\ Mod.\ Phys. D {\bf 3}, 529 (1994), gr-qc/9307018.

%%%%%%%%%%%%%%% Mikovic %%%%%%%%%%%%%%%
\bibitem{Mi1}
A.\ Mikovic, {\it Quantum gravity vacuum and invariants of embedded spin networks}, Class.\ Quant.\ Grav.\ {\bf 20}, 3483-3492 (2003), gr-qc/0301047.
\bibitem{Mi2}
A.\ Mikovic, {\it Flat spacetime vacuum in loop quantum gravity}, Class.\ Quant.\ Grav.\ {\bf 21}, 3909-3922 (2004), gr-qc/0404021.

%%%%%%%%%%%%%%% linearized gravity %%%%%%%%%%%%%%%
\bibitem{ARSgravitonsloops} A.\ Ashtekar, C.\ Rovelli, L.\ Smolin, {\it Gravitons and loops}, 
Phys.\ Rev.\ {\bf D44}, 1740-1755 (1991), hep-th/9202054.
\bibitem{V} M.\ Varadarajan, {\it Gravitons from a loop representation of linearized gravity}, 
Phys.\ Rev.\ {\bf D66}, 024017 (2002), gr-qc/0204067.

%%%%%%%%%%%%%%% coherent states %%%%%%%%%%%%%%%

\bibitem{TGCS1} 
 T.\ Thiemann, {\it Gauge field theory coherent states (GCS) : I.\ General properties}, Class.\ Quant.\ Grav.\ {\bf 18}, 2025-2064 (2001), hep-th/0005233.
\bibitem{TWGCS2}
 T.\ Thiemann, O.\ Winkler, {\it Gauge field theory coherent states (GCS): II.\ Peakedness properties}, Class.\ Quant.\ Grav.\ {\bf 18}, 2561-2636 (2001), hep-th/0005237.
\bibitem{TWGCS3}
 T.\ Thiemann, O.\ Winkler, {\it Gauge field theory coherent states (GCS): III.\ Ehrenfest theorems}, Class.\ Quant.\ Grav.\ {\bf 18}, 4629-4682 (2001), hep-th/0005234.
\bibitem{TWGCS4}
 T.\ Thiemann, O.\ Winkler, {\it Gauge field theory coherent states (GCS): IV.\ Infinite Tensor Product and Thermodynamical Limit}, Class.\ Quant.\ Grav.\ {\bf 18}, 4997-5054 (2001), hep-th/0005235.
\bibitem{SaTW}
H.\ Sahlmann, T.\ Thiemann, O.\ Winkler, {\it Coherent states for canonical quantum general relativity and the infinite tensor product extension}, Nucl.\ Phys.\ {\bf B606}, 401-440 (2001), gr-qc/0102038. 
\bibitem{Treviewcoherentstates}
T.\ Thiemann, {\it Complexifier coherent states for quantum general relativity}, gr-qc/0206037.
\bibitem{SaT1}
H.\ Sahlmann, T.\ Thiemann, {\it Towards the QFT on curved space-time limit of QGR.\ 1.\ General scheme}, gr-qc/0207030. 
\bibitem{SaT2}
H.\ Sahlmann, T.\ Thiemann, {\it Towards the QFT on curved space-time limit of QGR.\ 2.\ A concrete implementation}, gr-qc/0207031. 
\bibitem{AL}
A.\ Ashtekar, J.\ Lewandowski, {\it Relation between polymer and Fock excitations}, Class.\ Quant.\ Grav.\ {\bf 18}, L117-128 (2001), gr-qc/0107043.

%%%%%%%%%%%%%%% weaves %%%%%%%%%%%%%%%
\bibitem{ARSweaves} A.\ Ashtekar, C.\ Rovelli, L.\ Smolin, {\it Weaving a classical geometry with quantum threads}, Phys.\ Rev.\ Lett.\ {\bf 69}, 237-240 (1992), hep-th/9203079.
\bibitem{Ze} J.\ Zegwaard, {\it The weaving of curved geometries}, Phys.\ Lett.\ {\it B300}, 217-222 (1993), hep-th/9210033.
\bibitem{Bo} R.\ Borissov, {\it Weave states for plane gravitational waves}, Phys.\ Rev.\ {\bf D49}, 923-929 (1994).
\bibitem{Iw} J.\ Iwasaki, {\it Woven geometries: black holes}, gr-qc/9506028.
\bibitem{CRe} A.\ Corichi, A.M.\ Reyes, {\it A gaussian weave for kinematical loop quantum gravity}, 
Int.\ J.\ Mod.\ Phys.\ {\bf D10}, 325-338 (2001), gr-qc/0006067.

%%%%%%%%%%%%%%% general boundaries %%%%%%%%%%%%%%%
\bibitem{CDORT} F.\ Conrady, L.\ Doplicher, R.\ Oeckl, C.\ Rovelli,
M.\ Testa, {\it Minkowski vacuum in background independent quantum
gravity}, Phys.\ Rev.\ {\bf D69}, 064019 (2004), gr-qc/0307118.

%%%%%%%%%%%%%%% renormalization flow equations %%%%%%%%%%%%%%%

\bibitem{Reu}
M.\ Reuter, {\it Nonperturbative evolution equation for quantum gravity}, Phys.\ Rev.\ {\bf D57}, 971-985 (1998), hep-th/9605030.
\bibitem{ReuLau1}
O.\ Lauscher, M.\ Reuter, {\it Is quantum Einstein gravity nonperturbatively renormalizable?}, Class.\ Quant.\ Grav.\ {\bf 19}, 483-492 (2002), hep-th/0110021.
\bibitem{ReuLau2}
O.\ Lauscher, M.\ Reuter, {\it Is quantum Einstein gravity nonperturbatively renormalizable?}, Phys.\ Rev.\ {\bf D65}, 025013 (2002), hep-th/0108040.
\bibitem{PercPeri}
R.\ Percacci, D.\ Perini, {\it Asymptotic safety of gravity coupled to matter}, Phys.\ Rev.\ {\bf D68}, 044018 (2003), hep-th/0304222.

%%%%%%%%%%%%%%% renormalization %%%%%%%%%%%%%%%
\bibitem{Ma}
F.\ Markopoulou, {\it Coarse graining in spin foam models}, Class.\ Quant.\ Grav.\ {\bf 20}, 777-800 (2003), gr-qc/0203036.
\bibitem{O}
R.\ Oeckl, {\it Renormalization of discrete models without background}, Nucl.\ Phys.\ {\bf B657}, 107-138 (2003), gr-qc/0212047. 

\bibitem{Sstring} 
L.\ Smolin, {\it Strings as perturbations of evolving spin networks}, 
Nucl.\ Phys.\ Proc.\ Suppl.\ {\bf 88}, 103-113 (2000), hep-th/9801022.
\bibitem{MaScausalevolution} 
F.\ Markopoulou, L.\ Smolin, {\it Causal evolution of spin networks}, 
Nucl.\ Phys.\ {\bf B508}, 409-430 (1997), gr-qc/9702025.
\bibitem{MaSlocalcausality} 
F.\ Markopoulou, L.\ Smolin, {\it Quantum geometry with intrinsic local causality}, 
Phys.\ Rev.\ {\bf D58}, 084032 (1998), gr-qc/9712067.

%%%%%%%%%%%%%%% linearization in classical theory %%%%%%%%%%%%%%%
\bibitem{ALe} A.\ Ashtekar, J.\ Lee, {\it Weak field limit of general relativity in terms of new variables: a Hamiltonian framework}, Int.\ J.\ Mod.\ Phys.\ {\bf D3}, 675-693 (1994).
\bibitem{Hi}
A.\ Higuchi, {\it Linearized quantum gravity in flat space with toroidal topology}, Class.\ Quant.\ Grav.\ {\bf 8}, 2023-2034 (1991).

%%%%%%%%%%%%%%% others %%%%%%%%%%%%%%%
%\bibitem{VZ}
%M.\ Varadarajan, A.J.\ Zapata, {\it A proposal for analyzing the classical limit of kinematic loop gravity},
%Class.\ Quant.\ Grav.\ {\bf 17}, 4085-4110 (2000), gr-qc/0001040.
%\bibitem{Ar}
%M.\ Arnsdorf, {\it Loop quantum gravity and asymptotically flat spaces}, gr-qc/0008038.

\end{thebibliography}
\end{document}